\documentclass[12pt]{emulateapj}
\pdfoutput=1
\usepackage[english]{babel}
\usepackage{amsmath}
\usepackage{amssymb}
\usepackage{url}
\usepackage{color}
\usepackage{natbib}
\usepackage{bm}
\usepackage{mathtools}
\usepackage[flushleft]{threeparttable}

\usepackage[hypertexnames=false]{hyperref} 

\newcommand{\be}{\begin{equation}}
\newcommand{\ee}{\end{equation}}
\newcommand{\msun}{\,M_{\odot}}
\newcommand{\ergs}{\rm\,erg\,s^{-1}}
\newcommand{\beq}{\begin{eqnarray}}
\newcommand{\eeq}{\end{eqnarray}}

\begin{document}

\title{A Semi-Coherent Search for Weak Pulsations in Aql X--1}
\author{C. Messenger\altaffilmark{1}, A. Patruno\altaffilmark{2,3}} 
\altaffiltext{1}{SUPA, School of Physics and Astronomy, University of Glasgow, Glasgow, G12 8QQ, United Kingdom}
\altaffiltext{2}{Leiden Observatory, Leiden University, PO Box 9513, NL-2300 RA Leiden, the Netherlands}
\altaffiltext{3}{ASTRON, the Netherlands Institute for Radio Astronomy, Postbus 2, 7990 AA Dwingeloo, the Netherlands}

\begin{abstract}
\noindent
Non pulsating neutron stars in low mass X-ray binaries largely
outnumber those that show pulsations. The lack of detectable pulses
represents a big open problem for two important reasons. The first is
that the structure of the accretion flow in the region closest to the
neutron star is not well understood and it is therefore unclear what
is the mechanism that prevents the pulse formation. The second is that
the detection of pulsations would immediately reveal the spin of the
neutron star. Aql X--1 is a special source among low mass X-ray
binaries because it has showed the unique property of pulsating for
only ${\sim}$150 seconds out of a total observing time of more than 1.5
million seconds. However, the existing upper limits on the pulsed
fraction leave open two alternatives. Either Aql X--1 has very weak
pulses which have been undetected, or it has genuinely pulsed only for
a tiny amount of the observed time. Understanding which of the two
scenarios is the correct one is fundamental to increase our knowledge
about the pulse formation process and understand the chances we have
to detect weak pulses in other LMXBs. In this paper we perform a
semi-coherent search on the entire X-ray data available for Aql
X--1. We find no evidence for (new) weak pulsations with the most
stringent upper limits being of the order of 0.3\% in the 7--25 keV
energy band.
\end{abstract}
\maketitle

\section{Introduction}\label{sec:introduction}

Some neutron stars in low mass X-ray binaries (LMXBs) have
sufficiently strong magnetic fields to truncate the accretion disk and
channel the plasma along the field lines. According to accretion
theory~\citep{gho77,gho78, gho79}, the neutron star might be spun up
in the process with the gas impacting on the surface and forming
``hot spots'' plus a shock right above it
~\citep{bas76,pou03}. Modulation of the thermal and comptonized
radiation emerging in the process creates X-ray pulsations that reveal
the spin period of the neutron star. Accreting millisecond X-ray
pulsars (AMXPs) are neutron stars in low mass X-ray binaries (LMXBs)
with spin period of less than ${\sim}10$ ms, which are powered via the
process described above (\citealt{wij98}, see~\citealt{pat12r} for a
recent review).  However, among the ${\sim}150$ neutron stars in LMXBs,
only ${\sim}20$ show pulsations and have been unambiguously identified as
either AMXPs or as slow accreting pulsars (like Her X-1, GX 1+4, etc.,
see e.g., Table 1 in \citealt{pat12r} and \citealt{bil97}). The large
majority of neutron star LMXBs do not show accretion powered
pulsations with typical upper limits on the pulsed fraction in the
range of ${\sim}1$--$10\%$ rms (see for
example~\citealt{vau94,dib05}).

Many different possibilities have been proposed in the literature to
explain the paucity of pulsators among LMXBs, the most popular
including the onset of interchange instabilities that create a chaotic
accretion flow stream \citep{kul08}, the smearing and scattering of
pulsed emission \citep{bra87, tit02}, gravitational
lensing~\citep{woo88, oze09}, the nearly perfect alignment of the
neutron star magnetic and spin axis~\citep{rud91,lam09} and the
screening of the magnetosphere by the accreted matter~\citep{bis74,
  cum01}. So far, however, the exact reason behind this behaviour
remains not completely understood.

The discovery of the new phenomenon of
intermittent pulsations\citep{gal07} might help to shed light on the mechanism that
prevents most LMXBs from pulsating. Intermittent pulsations can be
described as a sporadic appearance and disappearance of X-ray pulses
(on variable timescales) during an outburst. 
HETE J1900.1--2455 was the first intermittent pulsar to be discovered
\citep{kaa06, gal07} and shows continuous pulsations for about 70 days
since the beginning of its nine year long outburst (still ongoing at
the moment of writing) with intermittent pulsations then appearing
sporadically for the next 2.5 years~\citep{gal07, gal08, pat12c}.
Pulsations were detected down to the 0.3\% rms level~\citep{gal08,
  pat12c}, whereas in the proceeding years pulsations were not
detected with the best upper limits on the fractional amplitude of
0.05\% rms (95\% confidence level, see~\citealt{pat12c}). During this
time HETE J1900.1--2455 was completely indistinguishable from one of
the many non-pulsating LMXBs (see e.g.,~\citealt{pap13}).
\citet{gav07} and \citet{alt08} discovered intermittent
pulses in SAX J1748.9--2021 that appeared sporadically throughout two
outbursts at unpredictable times.
Finally, \citet{cas08} made the particularly surprising discovery of a
single episode of pulsations in Aql X--1 (with 2--60 keV fractional
semi-amplitude\footnote{semi-amplitudes are $\sqrt{2}$ larger than
  rms-amplitudes} of approximately 2\%), that lasted for only $\approx
150$ s over a total observing time of about 1.5 Ms. This discovery
raised the question on whether all non pulsating LMXBs do show
pulsations for very brief time intervals that could be easily missed,
since the duty cycle of pulsations might be as short as in Aql X--1.

So far a big limitation in pulse searches has been the very large
computational time required to analyze the huge amount of data
available. The reason for this is that in most non-pulsating LMXBs the
orbital parameters of the system are poorly known, leaving a parameter
space too large to be searched. Indeed, simple techniques based on
power spectral density estimation are very limited in terms of
sensitivity if the pulse frequency drift due to Doppler motion in the
binary is not corrected for.  \citet{cas08} performed a complete
{\it{RXTE}} archival data search for pulsations in Aql X--1, using
standard Fourier transforms of 128s length. Such short data segments
ensured that the spin frequency stays in one--two Fourier frequency
bins during the observation, avoiding the spread of power in multiple
bins due to orbital motion Doppler shifts.  This search however is not
optimal, since the signal can be accumulated only in very short data
segments.

Another strategy employed is the so-called acceleration search
method~\citep{ran02}. This strategy requires a sub-division of data
into segments of no more than about 1/10 of the orbital period length
so to have an approximately \textit{constant} orbital acceleration
over each specific data segment. The signal is then searched by
summing the power in a certain amount of adjacent Fourier frequency
bins where the signal has spread due to the acceleration of the
neutron star.

The discovery of pulsations in all 17 known AMXPs (both intermittent
and persistent) can be ascribed to the use of the first method whereas
only the ultra-compact LMXB 4U 1820--30 has been thoroughly
searched with the acceleration technique leading to upper limits of
about 0.8\% rms on the pulsed fraction~\citep{dib05}.


In this paper we use a different approach to the problem.  To account
for the Doppler shift in the binary we use a so-called semi-coherent
search strategy, initially developed to optimize computationally
intensive gravitational wave searches~\citep{mes11} but implemented
and optimized in this work for deep pulse searches in X-ray
binaries. The concept of the search is a generalization and extension
of the acceleration search. Each segment of data is processed over a
bank of signal model template waveforms. The waveforms approximate the
binary Doppler modulation as a smooth phase evolution modelled by a
Taylor expansion in frequency. The data products from each segment are
represented by the Fourier power computed for each of these
templates. This power is then summed over segments such that the
excess power from all possible signals is tracked in frequency (and
frequency derivatives) as the source moves through its binary orbit.
This process affords an enhancement to the fractional amplitude
sensitivity approximately proportional to the fourth root of the number
of segments. Such a scheme is moderately computationally intensive and
requires the use of many 1000s of CPU hours.

We applied the semi-coherent search scheme to all archival data of Aql
X--1 recorded with high time resolution by the \textit{Rossi X-ray
  Timing Explorer} (\textit{RXTE}).  Since previous pulse searches
used short data segments of just 128 seconds, the sensitivity reached
was only sufficient to detect pulse fractional semi-amplitudes of the
order of 1\% (see \citealt{cas08}) which is very close to the reported
2--60 keV pulsed fraction (semi--amplitude) of the single pulse
episode ($1.9\pm0.2\%$).  Therefore it is plausible to expect that
what we have observed so far is not really a single intermittent
pulsation, which is indeed extremely problematic to explain from a
theoretical point of view, but only the ``tip of the iceberg'' with a
large amount of weak pulses lying below the detection sensitivity of
previous pulse searches. We therefore will direct our efforts towards
the search of weak (semi-amplitude of $\lesssim1\%$) but
\textit{continuous} pulsations.

In Secs.~\ref{sec:dataprep} and~\ref{sec:pspace} we discuss the \textit{RXTE}
data preparation and the Aql X--1 parameter space respectively. We then
describe the data preprocessing in Sec.~\ref{sec:preproc} and our
semi-coherent detection statistic in Sec.~\ref{sec:detstat}.  In
Sec.~\ref{sec:implement} we describe how our search for pulsations
from Aql X--1 was implemented and the corresponding results are
described in Sec.~\ref{sec:results}. A discussion of our findings is
given in Sec.~\ref{sec:discussion} and we conclude with
Sec.~\ref{sec:conclusions}.


\section{X-ray Observations and Data Preparation}\label{sec:dataprep}

\textit{RXTE} has observed Aql X--1 for ${\sim}15$ years, collecting data
of 20 outbursts and recording high time resolution data with the
Proportional Counter Array (PCA; see~\citealt{jah06}) for 18 of them. 
Each outburst lasts for a variable amount of time, from few days up to
six months, with a recurrence time of ${\sim}1$ yr (see
e.g.,~\citealt{cam13} and their Table 1).  We performed a complete
archival search on all {{\it RXTE}} public data available collected
between January 1997 and October 2010 (see Table~\ref{tab:obs}).

\begin{table}
\caption{{{\it RXTE}} observations of Aql X--1 from 1997 to 2010}
\centering
\scriptsize
\begin{tabular*}{\columnwidth}{@{\extracolsep{\fill}}lcr}
\hline
\hline
Year & Index & Program IDs \\
\hline
1997 Jan\dotfill & 1 & {\tt 20098}\\    
1997 Aug& 2 & {\tt 20091}\\
\smallskip
1998 Feb & 3 & {\tt 30072}, {\tt 30073}, {\tt 30188}\\
1999 May & 4 & {\tt 40033}, {\tt 40047}, {\tt 40048}
  {\tt 40049}, {\tt 40432}\\
2000 Sep& 5 & {\tt 50049}\\          
\smallskip
2001 Jun & 6 & {\tt 60054}\\
2002 Feb& 7 & {\tt 60429}, {\tt 70069}\\
2003 Feb& 8 &{\tt 70426}, {\tt 80403}\\
\smallskip
2004 Feb& 9 & {\tt 80403}, {\tt 90403},  {\tt 90017}\\
2005 Mar& 10 & {\tt 91028},  {\tt 91414}\\          
2005 Nov& 11 & {\tt 91414}\\
\smallskip
2006 Jul& 12 & {\tt 92034}\\
2007 May& 13 & {\tt 92438}, {\tt 92076} \\
2007 Sep& 14 & {\tt 93045}\\
\smallskip
2008 May& 15 & {\tt 93045}, {\tt 93076}\\          
2009 Mar& 16 & {\tt 94076}\\
2009 Nov& 17 & {\tt 94076}, {\tt 94441}\\
\smallskip
2010 Sep& 18 & {\tt 95086}, {\tt 95413}\\
\hline
\hline
\end{tabular*}
\label{tab:obs}
\end{table}

We used all pointed observations taken in GoodXenon or in Event
122$\mu\,s$ mode with time resolution of $2^{-20}$ s and $2^{-13}$ s,
respectively. The GoodXenon data were rebinned to match the same time
resolution of the Event data. We selected an energy band between 
7 and 25 keV which is based on the maximization
of the signal-to-noise ratio (S/N) of the single ${\sim}150$s pulse
episode previously detected in this source. Indeed, as noticed
by~\citet{cas08}, the pulsed fraction of Aql X--1 increases with
energy, growing from less than 1-2\% at low energies ($<5$ keV) to
10--20\% in the highest energy band (10--30 keV).

To inspect for the presence of thermonuclear bursts we construct the
2--16 keV X-ray light curve with the PCA Standard 2 data (16 s time
resolution). We refer to~\citet{van03} for further details on the
light-curve generation. The start and end time of thermonuclear bursts
are defined as the points where the count rate in the lightcurve is
twice the pre-burst value. The high time resolution data are then
barycentered at the best determined radio position of Aql
X--1~\citep{tud13} and are filtered according to standard procedures:
unstable pointings, thermonuclear bursts and passages through the
South Atlantic anomaly are removed from the data. When an X-ray burst
occurs, the data are split into two time-series, a pre-burst and a
post-burst. The largest majority of final-product time series have a
duration in the range 1--3 ks.

\section{Aql X--1: Parameter Space}\label{sec:pspace}

Aql X--1 has a relatively well constrained orbital and spin
parameters. The orbital period has been determined from optical
observations and constrained to be $18.9479\pm0.0002$
hr~\citep{che91,che98}. \citet{wel00} reported a slightly shorter
orbital period ($18.71\pm0.06$ hr) which was considered consistent
with the value reported by~\citet{che98} due to unaccounted
systematics.  We choose to define a safe orbital period range with
values between 18.5 and 19.2 hours. The orbital phase is considered
unknown and we therefore consider the range 0 to 1 cycles as our
search space.

The spin frequency is also known with good precision to be around 550
Hz thanks to burst oscillation measurements~\citep{zha98} and the
possible single accretion powered episode reported
by~\citet{cas08}. In particular,~\citet{cas08} reported a spin
frequency of $550.273(1)$ Hz, which is, however, not corrected for the
Doppler shift of the neutron star in the binary. To determine the
effect of the Doppler shift we explored a broad range of projected
semi-major axis values that span between 0.1 and 4.2 light-seconds.
In this case the term ``projected'' indicates the true orbital
semi-major axis projected along the line of sight of our observation.
Combined with our orbital periods, this gives a range of possible
pulse frequencies between 549.9 and 550.6 Hz.  Finally, we assume zero
orbital eccentricity, which is a good approximation for LMXBs. We note
that as indicated in Fig.~3 of~\citet{mes11} for our search and its
corresponding parameters we are insensitive to eccentricity below
$\lesssim 0.01$.

The complete physical Aql X--1 parameter space for this search is
therefore 4 dimensional. We assume no a-priori correlations between
our search parameters and hence our search space is equal to the
Cartesian product of the intrinsic spin frequency $f$, the projected
orbital semi-major axis $a$, the orbital period $P$ and the orbital
phase $\psi$.  This space is limited in each dimension by the ranges
specified above and in Table~\ref{tab:parspace}.
\begin{table}
  \caption{The pulse frequency and orbital parameter space boundaries for the AQL X--1 search.
    \label{tab:parspace}}
  \centering
  \begin{tabular*}{\columnwidth}{@{\extracolsep{\fill}}cccc}
    \hline \hline
    Parameter & Units & Min & Max \\ 
    \hline
    $f$ & Hz & 549.9 & 550.6\\
    $a$ & s & 0.1 & 4.2 \\
    $P$ & hr & 18.5 & 19.2 \\
    $\psi$ & rads & 0 & $2\pi$\\
    \hline \hline
  \end{tabular*}
\end{table}
%

\section{Data preprocessing}\label{sec:preproc}

The data is first divided into outbursts 1--18, and for each outburst
the data is comprised of multiple contiguous time-series.  Each
time-series is initially processed with a time-domain high pass 10th
order Butterworth filter with filter frequency $40$~Hz in order to
remove any spurious low-frequency modulation. Each contiguous stretch
of data is then further subdivided into segments of length $T=256$s.
This choice is based on computational constraints and is further
discussed in Sec.~\ref{sec:setup}. These ``segments'' now containing
gap-free time-series data are Fourier transformed according to
\begin{equation}\label{eq:FFTdef}
\tilde{x}_{k}=\sum_{j=0}^{N-1}x_{j}e^{-2\pi ijk/N}
\end{equation}
where the time-series data $x_{j}$ represents the photon count in the
$j$'th time bin and where the time index $j$ ranges from 0 to
$N=T/\Delta t$ and $\Delta t=2^{-13}$s is the sampling time.  Within
the process of subdividing into segments, stretches of data of length
$<T$ s or data left at the ends of time-series after division that were
$<T$ s were discarded at the expense of losing $\approx 5.7\%$ of the
total data.  Since our search concerns a relatively narrow frequency
band for the intrinsic spin frequency of the source, we also only
retain Fourier frequency bins within the range 549--552~Hz.

\section{The detection statistic}\label{sec:detstat}

Considering a single segment of X--ray data, we model our binned
timeseries $\bm{x}$ as Poisson distributed such that the likelihood function
for a single segment of data is
\begin{equation}\label{eq:likelihood}
  p(\bm{x}|\bm{\Theta},I) = \prod_{j=0}^{N-1}\frac{r_{j}(\bm{\Theta})^{x_{j}}e^{-r_{j}(\bm{\Theta})}}{x_{j}!}
\end{equation}
where $\bm{\Theta}$ is a vector of signal parameters (including our
search parameters) that define our signal model given by
\begin{equation}\label{eq:model}
  r_{j}(\bm{\Theta}) = R\left(1 + A\sin\left(\phi_{j}(\bm{\theta})+\beta\right)\right).
\end{equation}
Here $R$ is the expected background counts per time bin, $A$ is the
pulsed fraction of our signal, $\phi_{j}(\bm{\theta})$ is the time
dependent phase of the signal and $\beta$ is a reference phase of the
signal.  The parameter vector $\bm{\theta}$ are our search parameters
and a subset of the complete signal parameters $\bm{\Theta}$.

The null hypothesis assumes that no signal is present in the data and
we can therefore define the null, or noise-only, model using
Eq.~\ref{eq:model} and setting $A=0$.  The log-likelihood-ratio
between our signal and noise-only models can be approximated as
\begin{align}
  \Lambda &= \sum_{j=0}^{N-1} x_{j}\log\left(1 +
    A\sin\left(\phi_{j}(\bm{\theta})+\beta\right)\right)\nonumber \\
  &-RA\sin\left(\phi_{j}(\bm{\theta})+\beta\right)\nonumber\\
  &\approx -\frac{1}{4}NRA^{2} + A\sum_{j=0}^{N} x_{j}\sin\left(\phi_{j}(\bm{\theta})+\beta\right)
\end{align}
where we have assumed that we are working in the weak signal regime
and used the approximation $A\ll 1$.

The log-likelihood ratio can now be analytically maximized over the
unknown amplitude and phase parameters $A$ and $\beta$ to
give\footnote{We use $2\hat{\Lambda}$ here since the resulting
  statistic is then exactly $\chi^{2}$ distributed.}
\begin{equation}\label{eq:lambda}
  2\hat{\Lambda}(\bm{\theta})=2\frac{|\tilde{x}(\bm{\theta})|^2}{NR}
\end{equation}
where we define the phase model demodulated Fourier transform of the
data as
\begin{equation}\label{eq:FFTdemod}
  \tilde{x}(\bm{\theta})=\sum_{j=0}^{N-1}x_{j}\exp\left(-i\phi_{j}(\bm{\theta})\right).
\end{equation}
For a monochromatic phase model this expression is simply the discrete
Fourier transform evaluated at a specific frequency (equivalent to
Eq.~\ref{eq:FFTdef}).  In general our phase model will deviate from
monochromicity according to the binary motion of the source. This
statistic is $\chi^{2}$ distributed with 2 degrees of freedom and a
non-centrality parameter equal to the signal-to-noise ratio (SNR) of
the signal within this segment. Assuming a set of signal parameters
evaluated at an offset parameter space location
$\bm{\theta}+\bm{\Delta\theta}$, the expectation value of the single
segment detection statistic is
\begin{equation}
  \text{E}\left[2\hat{\Lambda}(\bm{\theta},\bm{\Delta\theta})\right]=2 +
  \rho^{2}(\bm{\theta},\bm{0})\left\vert\frac{1}{N}\sum\limits_{j=0}^{N-1}e^{-i\Delta\phi_{j}(\bm{\theta},\bm{\Delta\theta})}\right\vert^{2}.
\end{equation}
We have defined the optimal coherent SNR as the total noise-free
signal power weighted by the noise such that
\begin{equation}\label{eq:snr}
  \rho^{2}(\bm{\theta},\bm{0}) = \frac{4}{S_{\text{n}}}\int|\tilde{r}(\bm{\theta})|^2df=\frac{NRA^{2}}{2}
\end{equation}
where for all non-zero frequencies $S_{\text{n}}{=}2\Delta t R$
is the single-sided noise spectral density. Here we see the standard
result that the SNR is proportional to the signal amplitude and to the
square-root of the observation time since $T\propto N$.  

As outlined in~\cite{mes11} we aim to compute this statistic for each
segment and then sum the results over segments for many trial
$\bm{\theta}$ values.  We therefore define our semi-coherent statistic
as
\begin{equation}\label{eq:semico}
  \Sigma(\bm{\theta})=\sum_{m=1}^{M}2\Lambda_{m}(\bm{\theta})
\end{equation}
where $m$ indexes the segments ranging from 1 to $M$. For a dataset
containing a signal of pulse fraction $A$ and with phase model
parameters matching our template, the expectation value and variance
of our statistic is
\begin{subequations}
\begin{align}\label{eq:expsigma}
  \text{E}\left[\Sigma(\bm{\theta},\bm{0})\right] &= 2M +
  \frac{\mathcal{N}A^{2}}{2}\\
  \text{Var}\left[\Sigma(\bm{\theta},\bm{0})\right] &=4M + \mathcal{N}A^{2}
\end{align}
\end{subequations}
where we have used $\mathcal{N}=\sum_{m=1}^{M}NR_{k}$ to represent the
total number of photons accumulated during the entire observation. In
order to arrive at this expression we have used the properties
$\text{E}[x_{j}^{2}]=r_{j}(r_{j}+1)$ and
$\text{E}[x_{j}x_{l}]=r_{j}r_{l}$ for a Poisson distributed
variable. Based on Eq.~\ref{eq:expsigma} and our knowledge of its
underlying distribution we can directly interpret the second term,
dependent upon the signal amplitude, as the non-centrality parameter
of the $\chi^{2}$ distribution.

If we define the semi-coherent statistic SNR as the expected
difference in its value in units of the expected standard deviation
via
\begin{equation}
  \rho_{\Sigma}=\frac{\text{E}\left[\Sigma(\bm{\theta},\bm{0})-\Sigma(\bm{\theta},\bm{0};A=0)\right]}{\sqrt{\text{Var}\left[\Sigma(\bm{\theta},\bm{0})\right]}}
\end{equation}
we can broadly assess the sensitivity of our semi-coherent detection
statistic. For a fixed SNR we then find that in the weak signal limit
$A{\ll}1$ and $\mathcal{N}A^{2}{\ll}4M$ the amplitude satisfies
\begin{equation}\label{eq:sensitivity}
  A\propto\left(T\tau\right)^{-1/4}\equiv T^{-1/2}M^{-1/4}\equiv\tau^{-1/2}M^{1/4}.
\end{equation}
This is the standard result for semi-coherent searches.  For a fixed
total observation time the sensitivity to amplitude decreases as the
fourth-root of the number of segments.  A more rigorous calculation of
the search sensitivity is described in Secs.~\ref{sec:significance}
and~\ref{sec:upperlimits}.

\section{Implementation}\label{sec:implement}

Our analysis can be divided into 2 parts: the coherent demodulation of
signals within each of our data segments followed by the incoherent
combination of signal power from each segment.  In each case we use
banks of templates representing potential signal waveforms from within
our signal parameter space.  In the coherent stage of the analysis we
adopt a simplistic scheme for covering the parameter space and an
approximation to the full waveform model.  When combining the results
from each segment we return to the full waveform model but use a
template placement scheme based on randomly positioning templates in
the space.

\subsection{The coherent stage}\label{sec:coherent}

For our circular orbit model of the Aql X--1 system we define our phase
evolution as
\begin{equation}\label{eq:fullphase}
  \phi_{j}(\bm{\theta})=2\pi \nu\left[t-t_{0}+a\sin\left(\Omega (t-t_{0})-\gamma\right)\right]
\end{equation}
where $\nu$ is the intrinsic and constant spin frequency of the
neutron star, $a$ is the orbital radius projected along the line of
sight and normalized by the speed of light, $\Omega=2\pi/P$ is the
orbital angular frequency, $\gamma=\Omega(t_{0}-t_{\text{asc}})$ is an
orbital reference phase with $t_{\text{asc}}$ as the time of passage
through the ascending node and $t_{0}$ is a reference time.  We will
refer to this model in our discussion of the semi-coherent stage but
here at the coherent stage we choose to use a Taylor expanded
approximation to the phase model given by
\begin{equation}\label{eq:taylor}
  \phi_{j}(\bm{\nu}^{(m)}) = \phi_{0}^{(m)} + 2\pi \sum_{s=1}^{s*}\frac{\nu_{s}^{(m)}}{s!}(t_{j}-t_{\text{mid}}^{(m)})^{s}
\end{equation}
where
$\bm{\nu}^{(m)}=(\nu^{(m)}_{1},\nu^{(m)}_{2},\ldots,\nu^{(m)}_{s*})$
represents a vector of instantaneous frequency derivatives evaluated
at the mid-point of the $m$'th segment $t_{\text{mid}}^{(m)}$.  They
are defined as
\begin{equation}\label{eq:freqderivs}
  \nu_{s}^{(m)}=\nu a\Omega^{s}\sin\left(\gamma-\frac{s\pi}{2}\right)
\end{equation}
where we have chosen $t_{0}=t_{\text{mid}}$ for simplicity.  The
maximum number of derivatives to include in our approximate model
$s^{*}$ is defined prior to the search and chosen such that over the
length of a segment the maximal loss in recovered detection statistic
is below a predefined level. We discuss this in the next section.

We note that each set of $\bm{\nu}^{(m)}$ parameters are unique to
their specific data segment, i.e., a potential signal would be found
in each segment with different values of these parameters.  The
boundaries of this parameter space also change with each segment and
are identified as the frequency derivatives (Eq.~\ref{eq:freqderivs})
within each segment minimized and maximized over the range of possible
orbital parameters listed in Table~\ref{tab:parspace}. 

The computation of $|\tilde{x}(\bm{\nu}^{(m)})|^{2}$ is performed
efficiently via a resampling in the time-domain.  After Fourier
transforming the original time-series data $\bf{\tilde{x}}$, the
frequency region of interest is inverse Fourier transformed into a
real, down-sampled time-series according to
\begin{equation}
  \hat{x}^{(m)}_{j}=\frac{1}{N}\sum_{k=k^{*}}^{k^{*}+n-1}\tilde{x}^{(m)}_{k}e^{2\pi j(k-k^{*})/n}
\end{equation}
where $k^{*}$ and $n$ are the index of the lower bound on the
frequency region of interest and $n$ is the number of frequency bins in
that region respectively.  For each $\bm{\nu}^{(m)}$ template this new
timeseries is then resampled according to the time coordinate
\begin{equation}\label{eq:resample}
  \tau_{j}(\bm{\nu}^{(m)}) = \sum_{s=0}^{s*}\frac{\nu^{(m)}_{s}(t_{j}-t^{(m)}_{\text{mid}})^{s}}{\nu_{0}s!}.
\end{equation}
In order to obtain an arbitrary choice of frequency resolution in the
final stage of this process, zero-padding of the resampled timeseries
is also applied.  It is then finally transformed back to the frequency
domain via
\begin{equation}
  \tilde{x}_{k}(\bm{\nu}^{(m)})=\sum_{j=0}^{n-1}\hat{x}(\tau_{j})e^{-2\pi jk/n}
\end{equation}
to obtain the coherent detection statistic $\hat{\Lambda}(\bm{\nu}^{(m)})$.

\begin{figure}
  \begin{center}                                                                
    \includegraphics[width=\columnwidth]{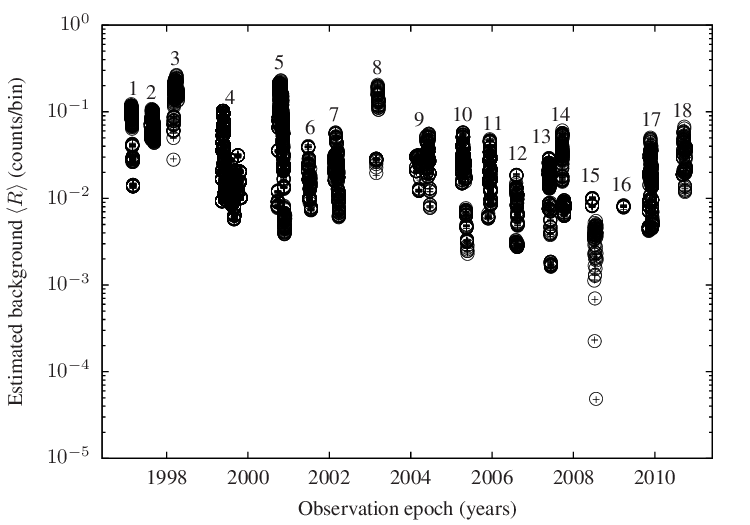}
  \end{center}                                                                  
  \caption{Estimates of the background counts per time bin ($\Delta
    t=2^{-13}$s) for each 256s segment used in the analysis. Black
    crosses indicate estimates computed via the median of the
    frequency domain power (Eq.~\ref{eq:noise}) and black circles
    correspond to estimates based on photon counts. The index
    of each outburst is also indicated.\label{fig:noise}}
\end{figure}

During the coherent stage the background count rate $R$ is estimated
from the data in each segment.  The actual value can be very precisely
obtained from counting the photons but in order to be robust against
deviation from the expected Poisson distribution of counts we use an
estimate obtained from the power spectrum according to
\begin{equation}\label{eq:noise}
  \langle R\rangle = \frac{\text{E}\left[|\tilde{x}|^2\right]}{N}.
\end{equation}
The expectation value of $|\tilde{x}|^2$ is estimated by computing the
discrete Fourier transform (without binary demodulation) of the
segment.  The median value is computed and converted to the mean
assuming a $\chi_{2}^{2}$ distribution and a median/mean ratio of
$\ln(2)$.  The values obtained from the frequency domain and photon
count methods are in excellent agreement and are shown in
Fig.~\ref{fig:noise} as a function of observation epoch. We note that
in practice the background estimation and subsequent normalization of
our coherent statistic $2\hat{\Lambda}$ is performed using a
running-median estimator of the frequency domain power spectrum to
remove any broad, non-pulselike, frequency domain features and the
background estimates shown in Fig.~\ref{fig:noise} are used in our
statistical significance and upper-limit calculations (see
Secs.~\ref{sec:significance} and~\ref{sec:upperlimits}).
%

\subsection{Coherent stage template placement}\label{sec:cotemplate}

We use a metric approach for template placement based on the expected
loss in SNR between a mismatched template and a
signal~\citep{bala96,owen96,owen99}.  The mismatch is a measure of the
fractional loss in squared SNR and can be approximated as
\begin{subequations}
\begin{align}
  \mu_{\Lambda}(\bm{\theta},\bm{\Delta\theta}) &= 1-\text{E}\left[\frac{\rho^{2}(\bm{\theta},\bm{\Delta\theta})}{\rho^{2}(\bm{\theta},\bm{0})}\right]\\
  &\approx
  -\frac{1}{2}\frac{1}{\rho^{2}(\bm{\theta},\bm{0})}\frac{\partial^{2}\rho^{2}(\bm{\theta})}{\partial\theta_{j}\partial\theta_{k}}\Delta\theta_{j}\Delta\theta_{k}\\
  &\approx g_{jk}(\bm{\theta}) \Delta\theta_{j}\Delta\theta_{k}
\end{align}
\end{subequations}
where $g_{jk}$ is the metric defined by
\begin{align}
  g_{jk}(\bm{\theta})=
  \left\langle\frac{\partial\phi(\bm{\theta})}{\partial\theta_{j}}\frac{\partial\phi(\bm{\theta})}{\partial\theta_{k}}\right\rangle
  - \left\langle\frac{\partial\phi(\bm{\theta})}{\partial\theta_{j}}\right\rangle \left\langle\frac{\partial\phi(\bm{\theta})}{\partial\theta_{k}}\right\rangle
\end{align}
and angled brackets represent the time average over the observation.
For our signal model, we have already analytically maximized over the
amplitude parameters and hence we are only concerned with mismatches
on the phase parameters $\bm{\theta}$.  It follows that the metric is
a function of derivatives of the phase model with respect to these
parameters.

For our approximate phase model used as defined in Eq.~\ref{eq:taylor}
we are able to compute the following metric
\begin{equation}
  g^{(\Lambda)}_{jk}(\bm{\nu})=\left(\begin{array}{cccc}
      \pi^2 T^{2}/3 & 0 & \pi^{2}T^{4}/120 & \ldots\\
      0 & \pi^2 T^4/180 & 0 & \ldots\\
      \pi^{2}T^{4}/120 & 0 & \pi^{2}T^{6}/4032 & \ldots\\
      \vdots & \vdots & \vdots & \ddots
\end{array}\right)
\end{equation}
from which we see that there are correlations (off-diagonal terms)
between some parameters.  For practical purposes we choose to take
only the diagonal terms leading to a conservative over-density of
templates. Also we note that our parameterization of the phase leads
to a ``flat'' metric where none of the elements are dependent upon any
of the phase parameters.  This leads to template spacing that remains
constant over the parameter space.

Templates, equivalent to locations within our space, are then
positioned such that any potential signal would incur a predefined
maximum mismatch in a worst case scenario.  The metric equates
distances between parameter space locations to this measure of
mismatch and, together with a gridding strategy, informs us on how to
place templates optimally.  In this case, optimally should be
interpreted as the minimum number of templates required to cover the
space given a maximally allowed mismatch. Using a hypercubic lattice
of parameter space locations and using only diagonal metric components
we can compute the spacing according to
\begin{equation}\label{eq:spacing}
  \Delta\nu_j=2\sqrt{\frac{\mu^{*}}{s^{*}g_{jj}}}.
\end{equation}
This guarantees that in a worst case scenario where a true signal has
parameter values that lie in the centre of hypercubic cell
(equidistant in mismatch from each of the closest templates) that the
total mismatch is maximally equal to $\mu^{*}$.  The output of
the coherent stage of the analysis is then the log-likelihood-ratio
$\Lambda$ (Eq.~\ref{eq:lambda}) computed on banks of templates on the
$\bm{\nu}^{(m)}$ parameter space for each segment.

In order to define the number of dimensions $s^{*}$ required to
accurately approximate the phase with our model, for each segment we
compute the number of templates that span the parameter space range.
This range is computed by finding the maximum span of
Eq.~\ref{eq:freqderivs} after varying the search parameters over their
respective ranges (given in Table~\ref{tab:parspace}).  This is done
with the exception of $\nu$ which is held fixed at its maximum value
within sub-bands over the frequency search space.

\subsection{The semi-coherent stage}\label{sec:semico}

The semi-coherent detection statistic $\Sigma(\bm{\theta})$, defined
in Eq.~\ref{eq:semico}, is the sum of individual coherent statistics
from each of the $M$ segments.  The corresponding semi-coherent
mismatch (defined as the loss of semi-coherently summed SNR) is then
\begin{subequations}
  \begin{align}
    \mu_{\Sigma}(\bm{\theta},\bm{\Delta\theta})&=
    1-\text{E}\left[\frac{1}{M}\sum\limits_{m=1}^{M}\frac{\rho_{m}^{2}(\bm{\theta},\bm{\Delta\theta})}{\rho_{m}^{2}(\bm{\theta},\bm{0})}\right]\\
    &=\frac{1}{M}\sum\limits_{1}^{M}\mu^{(m)}_{\Lambda}(\bm{\theta},\bm{\Delta\theta}).
  \end{align}
\end{subequations}
It follows that the metric defined on the semi-coherent mismatch is
simply the average of the individual segment coherent
metrics~\citep{bc00}. In the physical binary parameter space
$\bm{\theta}=(\nu,a,\gamma,\Omega)$ the semi-coherent metric has been
computed by~\citet{mes11} and is given by
\begin{equation}
   g^{(\Sigma)}_{jk}\approx\frac{\left(\pi T\right)^{2}}{6}\left(\begin{array}{cccc}
      2 & 0 & 0 & 0\\
      0 & \left(\nu\Omega\right)^2 & 0 & 0\\
      0 & 0 & \left(\nu a\Omega\right)^{2} & 0\\
      0 & 0 & 0 & \tfrac{1}{12}\left(\nu a\Omega\tau\right)^{2}
\end{array}\right)
\end{equation}
where $\Omega=2\pi/P$ is the orbital angular frequency. This metric is
specific to the case where $T{\ll}\tau$ and $T{\ll}P$ where $\tau$ is
the total observation span.  This is the case for our Aql X--1 search
where $P$ is ${\sim}19$ hours, $T=256$s and for each outburst $\tau$
is $\mathcal{O}($weeks--months$)$.  We also rely on the fact that the
segments are approximately evenly distributed over the entire orbital
cycle.

\subsection{Semi-coherent stage template placement}\label{sec:semitemplate}

For template placement at the semi-coherent stage we adopt the techniques
proposed in~\citet{mes11} and use a random template bank on the $\bm{\theta}$
parameter space. The semi-coherent metric is not constant across the range of
the parameter space and hence template spacings are variable on all parameters
with the exception of the frequency.  The fact that the metric is diagonal
allows us to perform a simple reparameterization to flatten the metric and would
allow us to use a lattice as opposed to a random covering.  However, for
simplicity a random covering was used where we first compute the number of
templates required via
\begin{align}\label{eq:nrandom}
n&=\log\left(\frac{1}{1-\eta}\right)\frac{\pi^{4}
T^{4}\tau}{25920m^2}\left(\nu_{\mathrm{max}}^{4}-\nu_{\mathrm{min}}^{4}\right)\nonumber\\
&\times\left(a_{\mathrm{max}}^{3}-a_{\mathrm{min}}^{3}\right)
\left(\Omega_{\mathrm{max}}^{4}-\Omega_{\mathrm{min}}^{4}\right)\left(\gamma_{\mathrm{max}}-\gamma_{\mathrm{min}}\right)
\end{align}
where $\mu$ is the desired nominal mismatch and $\eta$ is the covering
probability.  The covering probability is the probability of any particular
point in the space having a mismatch $<\mu$.  If we substitute the parameter
space ranges and search parameter choices for our Aql X--1 search we obtain the
following estimate,
\begin{equation}\label{eq:naqlx1} n \approx 2.3\times
10^{10}\left(\frac{T}{256\text{s}}\right)^4\left(\frac{\tau}{1\,\text{month}}\right)
\end{equation}
for a typical observation span and for $\mu=0.1$ and $\eta=0.9$.  The actual
number of templates used for each outburst analysis is given in
Table~\ref{tab:datapar}. 

Then $n$ points are randomly positioned on the physical
parameter space $\bm{\theta}$ with density $d$ proportional to the
square-root of the metric determinant such that 
\begin{equation}\label{eq:randomdensity}
d\propto \nu^{3}\Omega^{3}a^{2}
\end{equation}
Whilst this scheme, in 4-dimensions, results in ${\sim}30\%$ more
templates than the most optimal lattice placement strategy it contains
${<}1/2$ the number of templates of a basic hyper-cubic approach. On
average with a random template bank in 4 dimensions and with
$\eta=0.9$ the expected mismatch at any given point is ${\approx}60\%$
of the nominal mismatch value.  In our search this value was $\mu=0.1$
and hence on average we would expect a loss of $6\%$ in SNR from our
semi-coherent template placement strategy.

We have only defined the number of semi-coherent templates
since in all cases they far exceed the number of coherent templates and
dominate the computational cost. It is clear from Eq.~\ref{eq:naqlx1} exactly
how sensitive the number of templates is to our choice of coherent
observation length and consequently how we are constrained in this case to
using $T=256$s.

\subsection{Combining coherent results}\label{sec:combining}

Our semi-coherent detection statistic formulation in
Sec.~\ref{sec:detstat} implies that for every semi-coherent template
we compute the value of $\tilde{x}_{k}(\bm{\theta})$ for each segment.
In practice this is computationally prohibitive and instead, as
described above, we precompute this quantity for each segment on
hyper-cubic grids of the $\bm{\nu}^{(k)}$ parameters that define our
approximate templates.  When combining results by summing over
segments, for each semi-coherent template we compute the corresponding
instantaneous frequency derivatives at the midpoints of each segments
and then perform nearest neighbor interpolation on the precomputed
quantities $\tilde{x}_{k}(\bm{\nu})$.  Our semi-coherent statistic
then becomes
\begin{equation}\label{eq:semicoapp}
  \Sigma\left(\bm{\theta}\right)\approx\sum\limits_{m=1}^{M}2\Lambda\left(\bm{\nu}_{\text{nn}}^{(m)}(\bm{\theta})\right)
\end{equation}
where $\bm{\nu}_{\text{nn}}^{(m)}(\bm{\theta})$ is the nearest
neighbor location in $\bm{\nu}^{(m)}$ space in reference to the exact
location computed via Eq.~\ref{eq:freqderivs}.

In this approach, the hyper-cubic grids used on the $\bm{\nu}^{(m)}$
space highly simplify the interpolation procedure and counter-act the
cost of their original over-sampling (since hyper-cubic grids are not
the most efficient covering). The coherent templates are placed with a
maximal mismatch of $\mu=0.1$.  Since the relative location of a
potential signal with respect to the templates will vary between
segments, the summed statistics and their SNR losses will be subject
to averaging.  For the hyper-cubic grid in any number of dimensions
this results in an expected summed mismatch of $1/3$ the maximal
value. This loss in SNR is in addition to the losses incurred from the
mismatch in the semi-coherent template bank itself and for small
mismatches ${<}0.1$ can be assumed to additive.

\section{Results}\label{sec:results}

\begin{figure}
  \begin{center}                                                                
    \includegraphics[width=\columnwidth]{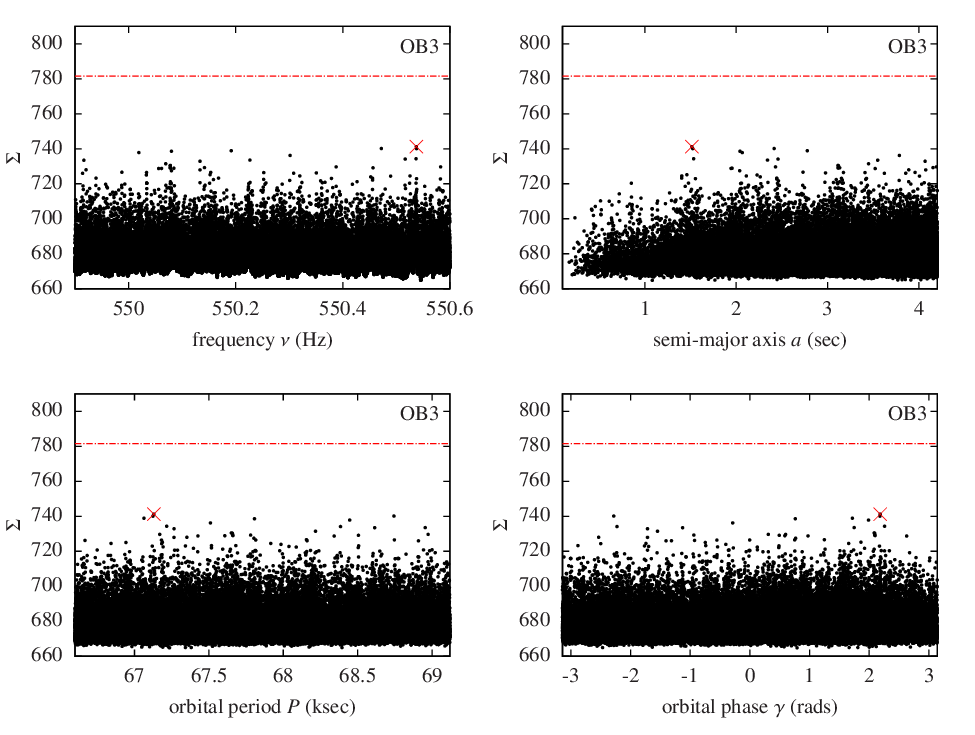}
  \end{center}                                                                  
  \caption{The semi-coherent detection statistic $\Sigma$ plotted as a
    function of the 4-dimensional physical search space for the 3rd
    Aql X--1 outburst. The red cross indicates the location and value
    of the loudest detection statistic and the dashed horizontal line
    indicates the $1\%$ false alarm threshold.The results plotted here
    are the 100 loudest statistics in each $1$mHz sub-band.\label{fig:ob3}}
\end{figure}

The application of our semi-coherent search to \textit{RXTE} observations of Aql
X--1 returned no evidence for the detection of pulsations in any of the 18
outbursts analysed.  The corresponding search parameters, maximum detection
statistics, and derived pulse fraction upper-limits are given in
Table~\ref{tab:datapar}. The derived upper-limits form the main result of the
analysis and in our most sensitive outburst we are able to limit the pulse
fraction to ${<}0.249\%$ with $90\%$ confidence with the majority of outbursts
return upper-limits ${<}1\%$.  Results of the search in the most sensitive
dataset (outburst 3\footnote{Outburst 3 is the dataset within which coherent
pulsations were originally detected in Aql X--1~\citet{cas08}. In our analysis
we have excluded the final 150 s of the outburst where these pulsations were
seen.}) are shown in Fig.~\ref{fig:ob3}. In the following sections we describe
the Aql X--1 search setup, the statistical analysis of the search results and
the derivation of upper-limits, and finally we present a validation of the
search using simulated signals.

\begin{table*}
  \centering
  \caption{The AQL X-1 outburst data parameters, estimated
    sensitivities and upper limits.
    \label{tab:datapar}}
  \begin{tabular}{cccccccccccc}
    \hline \hline
    Outburst & GPS start & $\tau$ Ms & $\mathcal{N}\times 10^{6}$ & $T$
    ks & $M_{256}$ &
    $n\times 10^{9}$ & $\Sigma^{1\%}$ & $A^{1\%}_{10\%}\,(\%)$ &
    $\Sigma^{*}$ & $P_{n}(\Sigma^{*})$ & $A^{\mathrm{UL90\%}}\,(\%)$ \\ 
    \hline
    1 & 540168821 & 1.478 & 46.8 & 77.06 & 301 & 13.04 & 880.9 & 0.396 & 846.6 & 0.8748 & 0.374\\
2 & 555367363 & 2.675 & 79.36 & 151 & 590 & 23.61 & 1562 & 0.352 & 1501 & 1.0 & 0.328\\
3 & 572925014 & 3.366 & 97.51 & 66.05 & 258 & 29.71 & 781.6 & 0.266 & 741.2 & 0.9996 & 0.249\\
4 & 610577860 & 14.47 & 78.44 & 371.2 & 1450 & 127.7 & 3498 & 0.441 & 3456 & 0.295 & 0.428\\
5 & 653812203 & 5.785 & 174.7 & 254.2 & 993 & 51.05 & 2478 & 0.269 & 2424 & 0.854 & 0.256\\
6 & 677384433 & 2.258 & 7.657 & 51.46 & 201 & 19.93 & 638.5 & 0.902 & 617.3 & 0.3949 & 0.867\\
7 & 697855525 & 3.46 & 18.75 & 103.7 & 405 & 30.53 & 1134 & 0.669 & 1106 & 0.4183 & 0.643\\
8 & 730230099 & 1.678 & 16.64 & 18.43 & 72 & 14.81 & 298.9 & 0.501 & 270.4 & 1.0 & 0.459\\
9 & 761167047 & 10.78 & 19.65 & 82.18 & 321 & 95.13 & 941.3 & 0.630 & 922.8 & 0.1728 & 0.611\\
10 & 796608483 & 4.422 & 20.41 & 93.7 & 366 & 39.02 & 1043 & 0.631 & 1006 & 0.9056 & 0.597\\
11 & 817073265 & 2.024 & 12.78 & 74.24 & 290 & 17.86 & 856.3 & 0.753 & 819.1 & 0.9703 & 0.705\\
12 & 838546705 & 1.389 & 2.698 & 38.66 & 151 & 12.26 & 509.4 & 1.426 & 476.5 & 0.999 & 1.328\\
13 & 863862099 & 1.756 & 8.983 & 68.35 & 267 & 15.5 & 799.7 & 0.881 & 769.4 & 0.7612 & 0.837\\
14 & 874369790 & 2.116 & 12.94 & 64.77 & 253 & 18.67 & 766.7 & 0.727 & 732.9 & 0.9374 & 0.685\\
15 & 897654239 & 3.321 & 0.9707 & 24.06 & 94 & 29.31 & 362.6 & 2.195 & 335.2 & 0.9986 & 2.038\\
16 & 921996024 & 0.007 & 0.186 & 2.816 & 11 & 0.0623 & 91.92 & 3.302 & 71.46 & 1.0 & 2.865\\
17 & 941531738 & 2.913 & 14.37 & 92.93 & 363 & 25.71 & 1033 & 0.746 & 992.1 & 0.9821 & 0.701\\
18 & 967988518 & 2.33 & 7.949 & 29.7 & 116 & 20.56 & 420.4 & 0.795 & 407.6 & 0.1649 & 0.770\\

    \hline \hline
  \end{tabular}  
  \begin{tablenotes}
    \footnotesize
  \item The column headings indicate the outburst index, the GPS start
    time of the outburst, the time span of the outburst observations,
    the total number of photons within the outburst, the total
    on-source data used, the number of 256s segments, the number of
    semi-coherent templates, the $1\%$ multi-trial false alarm
    threshold on $\Sigma$, the pulse fraction corresponding to a $1\%$
    multi-trial false alarm and a $10\%$ false dismissal probability,
    the loudest measured detection statistic $\Sigma^{*}$, the
    multi-trial statistical significance of $\Sigma^{*}$, and the
    $90\%$ confidence pulse fraction upper-limit.
  \end{tablenotes}
\end{table*}
\begin{figure}
  \begin{center}          
    \includegraphics[width=\columnwidth]{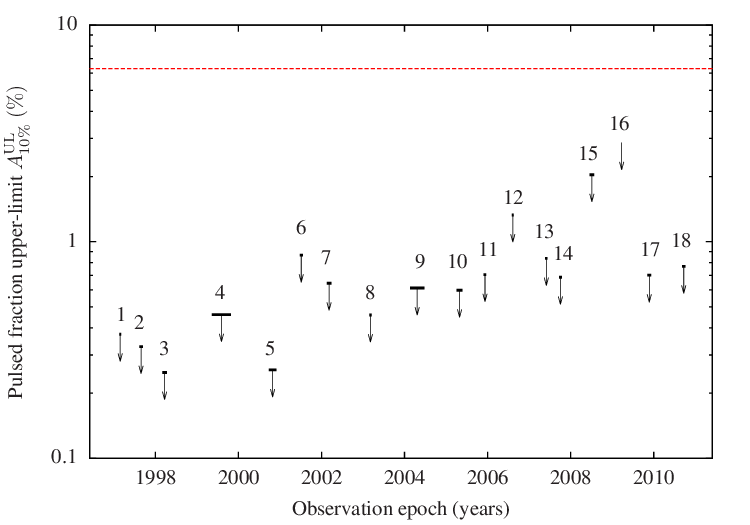}
  \end{center}                                                                  
  \caption{The $90\%$ confidence upper-limit on the pulse fraction for
    all outbursts as a function of observation.  For each outburst the
    span of the solid horizontal bar represents the span of the
    observation.  The dashed horizontal line indicates the pulse
    fraction observed by~\citep{cas08} and we have also indexed each
    outburst.\label{fig:ULs}}
\end{figure}  

As can be seen from Fig.~\ref{fig:ob3} and in the results from the
other outbursts, there is a general trend for the detection statistic
to be uniformly distributed with respect to the frequency, orbital
period and orbital phase parameters.  There is an clear increase in
the occurrence of larger values of the statistic at higher values of
the semi-major axis.  This is to be expected since the density of
templates also increases with semi-major axis.  Hence, per unit
semi-major axis there is a higher trials factor and correspondingly higher
expectation in the loudest statistics recorded.  We also note
that only a subset of the $29.71\times 10^9$ results are plotted for
outburst 3.  The analysis is split into 700 separate $1$mHz sub-bands
for processing in parallel using the
ATLAS\footnote{https://wiki.atlas.aei.uni-hannover.de} computer
cluster~\citep{aul09} and in Fig.~\ref{fig:ob3} we plot only the
loudest 100 statistics per sub-band. Finally, we note that the reference time
$t_{0}$ used to define the orbital phase $\gamma$ (see Sec.~\ref{sec:coherent})
is equal to the mid point of the observation span for each outburst. 

For all outbursts a threshold value on the semi-coherent statistic is
determined corresponding to a conservative approximation to a $1\%$ false alarm
rate.  In all cases no statistic exceeded this threshold and hence all results,
even those exhibiting peak-like structures, were consistent with the null
hypothesis.  Peak-like structures are a natural feature of the noise (as
verified in our simulations) and since our templates are, by design, highly
correlated, any loud statistic values will be locally surrounded by similarly
loud values.  As an additional check we have performed a search on our most
sensitive outburst (OB3) with a greatly extended orbital period parameter space
ranging from 5 to 20 hours.  Such short orbital periods increase the
computational cost of the search by large factors (see Eq.~\ref{eq:naqlx1}) and
to counteract this effect the coherent observation length was reduced to 32s.
In this case, via Eq.~\ref{eq:sensitivity} this corresponds to a reduction in
the search sensitivity by $\sim\sqrt{2}$ . No statistically significant
detection statistic values were recorded.

\subsection{Aql X--1 search setup}\label{sec:setup}

Our choice of coherent observation time $T=256$ s was motivated by
computational limitations specifically in the number of semi-coherent
templates.  From Eq.~\ref{eq:nrandom} we see that the number of
templates is proportional to $T^4$ however, as we will show, the
sensitivity of semi-coherent searches to the pulse fraction $A$
is proportional to $T^{1/4}$ (for a fixed total observation length).
Hence, our value of $T$ has been chosen so as to achieve near optimal
sensitivity while also keeping analysis times at manageable levels.
Other freely chosen parameters of the analysis were the coherent and
semi-coherent template bank mismatches which were both set to
$\mu=0.1$.  For the coherent template bank this represents the worst
case mismatch and has a corresponding average value of $0.03$.  For
the semi-coherent case $\mu=0.1$ represents the mismatch achieved with
a coverage probability $\eta=0.9$.  The resulting average mismatch at
any given parameter space location is $0.06$. Finally, for $s^{*}$,
the maximum number of search dimensions on the approximate phase
model, Eq.~\ref{eq:spacing} was used to determine the required
template spacing and then compared to the maximal parameter space
width in the corresponding dimension. If the spacing was greater than
the width then the dimension was not considered as part of the phase
model. For our choice of mismatch and coherent observation time
together with the Aql X--1 parameter space this resulted in a maximum
value of $s^{*}=2$. 

The \textit{RXTE} observations of Aql X--1 span ${\sim}13$ years and are divided
into 18 outbursts which we have chosen
to analyse separately.  The typical time span of an outburst is
$\mathcal{O}(\text{few})$ Ms and together with our additional search
parameter choices makes each analysis computationally tractable over a
the timescale of ${\sim}$days using ${\sim}10^3$ nodes of a the ATLAS
computer cluster. A single analysis of the entire dataset is made
computationally very difficult by the linear relationship between the
number of semi-coherent templates and the total observation span
$\tau$.  Such an analysis using the same coherent observation length
would therefore be ${\sim}100$ times more intensive with a gain of
only ${\approx}2$ in sensitivity to pulse fraction. Since our search
is sensitive to signals of duration equal to or greater than our total
observation, our choice of subdivision of analyses increases our
sensitivity to signals of duration ${\ge}1$ Msec.

\subsection{Statistical significance}\label{sec:significance}

A common problem in a templated wide parameter space search is the
difficulty in estimating the number of templates that constitute
statistically independent trials.  By design we aim to have highly
correlated templates, placed closely enough so as to not miss
potential signals.  By taking the actual number of templates as an
upper-limit on the number of trials we can compute correspondingly
conservative lower-limits on detection significance.  Let the
probability of obtaining a value of our statistic greater than or
equal to $\Sigma$ be $P(\Sigma)$ in the case of noise alone and a
single trial.  The distribution of $\Sigma$ for noise only is known to
be the central $\chi^{2}$ distribution with $2M$ degrees of freedom
and hence 
\begin{equation}\label{eq:P1}
  P(\Sigma) = 1 - \frac{\gamma\left(M,\frac{\Sigma}{2}\right)}{\Gamma(M)}
\end{equation}
where $\gamma$ and $\Gamma$ represent incomplete and complete gamma
functions respectively.  With an upper-limit on the number of
independent trials equal to $n$ we can state that
\begin{equation}\label{eq:PnofSigma}
  P_{n}(\Sigma) \leq 1 - \left(1-P(\Sigma)\right)^{n}
\end{equation}
is the probability of getting 1 or more events greater than $\Sigma$
after $n$ trials. We can now equate this to a multi-trial false alarm
probability $P_{\text{fa}}$ and solve for $\Sigma$ giving
\begin{equation}\label{eq:Sigmathresh}
  \Sigma^{*}\leq P^{-1}\left((1-P_{\text{fa}})^{1/n}-1\right)
\end{equation}
where $P^{-1}$ is the inverse function of the single trial
probability.  We show in Table~\ref{tab:datapar} the outburst
parameters and the expected sensitivities to pulse fraction amplitude
for a multi-trial false alarm of $1\%$ and a false dismissal
probability of $10\%$.  We also give upper-limits on the statistical
significance of the loudest events in each outburst.

\subsection{Upper-limits on pulse fraction}\label{sec:upperlimits}

Given the results of our analyses are consistent with the null
hypothesis we proceed to set upper-limits on the pulse fraction in
each outburst. We base this on our loudest statistic and ask the
question ``what is the value of $A$ such that with probability $C$ we
would have achieved a detection statistic greater than, or equal to,
the maximum value observed $\Sigma^{*}$''.  Using the expected
distribution of the detection statistic in the presence of a signal
(the properties of which are given in Eq.~\ref{eq:expsigma}) we solve
the following for $A$:
\begin{align}\label{eq:UL}
  C(\Sigma^{*},\mu^{*}) =&
  \int\limits_{\Sigma^{*}}^{\infty}d\Sigma\int\limits_{0}^{1}d\mu_{\Sigma}\int\limits_{0}^{\mu^{*}}d\mu^{(1)}_{\Lambda},\ldots \int\limits_{0}^{\mu^{*}}d\mu^{(M)}_{\Lambda}\nonumber\\
  &\chi^{2}_{2M}\left[\Sigma,\lambda(A,\mu_{\Sigma},\left\{\mu_{\Lambda}\right\})\right]p(\mu_{\Sigma})\prod_{m=1}^{M}p(\mu_{\Lambda}^{(m)}).
\end{align}
where $\mu^{*}$ is the maximal coherent template mismatch and
$p(\mu_{\Sigma})$ and $p(\mu_{\Lambda}^{(m)})$ are the prior mismatch
distributions for the semi-coherent and coherent template banks
respectively. The non-centrality parameter $\lambda$ of the
non-central $\chi^{2}$ likelihood function is simply the sum of
squared SNRs from each segment after accounting for mismatches such
that
\begin{equation}\label{eq:lambdaUL}
  \lambda(A,\mu_{\Lambda},\mu_{\Sigma})=\frac{NA^{2}}{2}(1-\mu_{\Sigma})\sum_{m=1}^{M}\langle
  R\rangle_{m}(1-\mu^{(m)}_{\Lambda}).
\end{equation}
We marginalize the likelihood over the possible mismatches expected
from both the coherent and semi-coherent template banks. This
expression is an accurate approximation despite the fact that we have
assumed the same average background rate for each segment. For the
hypercubic grid of coherent templates we know that the probability
distribution on mismatch for a single random location in 2-dimensions
is given by
\begin{equation}\label{eq:pmulambda}
p(\mu_{\Lambda})=
\begin{dcases*}      
   \pi/2 & if
    $\mu\leq 1/2$ \\
   \frac{\pi}{2} - 2\cos^{-1}\left(\frac{1}{\sqrt{2\mu_{\Lambda}}}\right)
    & if $1/2<\mu<1$.
  \end{dcases*}
\end{equation}
For the semi-coherent bank a single statistic is affected by only one
realization of mismatch and in 4-dimensions is governed by the
distribution
\begin{equation}\label{eq:pmusigma}
  p(\mu_{\Sigma},\eta)=-2\log\left(1-\eta\right)\frac{\mu_{\Sigma}}{\mu^{*}}(1-\eta)e^{-(\mu_{\Sigma}/\mu^{*})^2}.
\end{equation}

Using Eq.~\ref{eq:UL} we are then able to answer our upper-limit
question and claim an amplitude on pulse fraction above which we
confidently rule out the true signal value.  The corresponding values
for each Aql X--1 outburst are given in Table~\ref{tab:datapar} in the
final column. We can also use Eq.~\ref{eq:UL} to compute an expected
search sensitivity based on a predefined false alarm and false
dismissal probability. In this case we simply replace the input
measured semi-coherent statistic with the value of
$\Sigma^{P_{\text{fa}}}$ computed via Eq.~\ref{eq:Sigmathresh} and
equate our upper-limit confidence $C$ to the complement of the false
dismissal probability.  The corresponding values of $A$ for a false
alarm probability of $1\%$ and a false dismissal of $10\%$ ($C=90\%$)
are also given Table~\ref{tab:datapar}.

\subsection{Pulse Search Validation}\label{sec:validation}

To verify our claimed sensitivity (pulse fraction semi-amplitude of
$0.3\%$) and to validate our search algorithm, we performed a blind
test in which an artificial signal of amplitude close to the claimed
sensitivity was injected into a random outburst. The binary and spin
parameters of the signal were randomly chosen by one author from
within the range reported in Table~\ref{tab:parspace} and with a
semi-amplitude of $0.4\%$.  The same author then replaced one of the
first six outbursts\footnote{we chose to use only six outbursts
  instead of eighteen to save computational power} with an artificial
outburst containing the signal while maintaining the statistical
properties of the original outburst.
Then the first six outbursts (including the artificial
one) were submitted to the other author who, without knowing which
outburst and which binary/spin parameters were selected, proceeded to
apply the search algorithm to the datasets. The results show that the
outburst containing the fake signal (outburst 4) was detected with a
false alarm probability of $\leq 3.6\times 10^{-5}$ and would have
therefore been claimed as a detection.

We are able to make relatively accurate estimates of the parameter
uncertainties using Bayes theorem together with some simplifying
assumptions. In the specific case where the prior probability
distributions on the search parameters are uniform we find that the
posterior distribution on the search parameters is proportional to the
likelihood function.

The first of our simplifications is to use only the loudest template
to perform any inference and to treat the pulse fraction separately
from the phase parameters.  Our estimate of the true pulse fraction
value is therefore obtained according to
\begin{equation}\label{eq:postA}
  p(A|\Sigma^{*}) \propto \int\limits_{0}^{\infty}d\mu_{\Sigma}\chi^{2}_{2M}\left(\Sigma^{*},\lambda\left(A,\{\mu_{\Lambda}\},\mu_{\Sigma}\right)\right)p(\mu_{\Sigma})
\end{equation}
where we include a marginalization over the unknown value of the true
semi-coherent mismatch.  From this posterior we then take the median
as our pulse fraction estimate $A^{*}$ and compute the minimal $68\%$
confidence range as our uncertainty (see Table~\ref{tab:sigpars}).

\begin{figure}
  \begin{center}                                                                
    \includegraphics[width=\columnwidth]{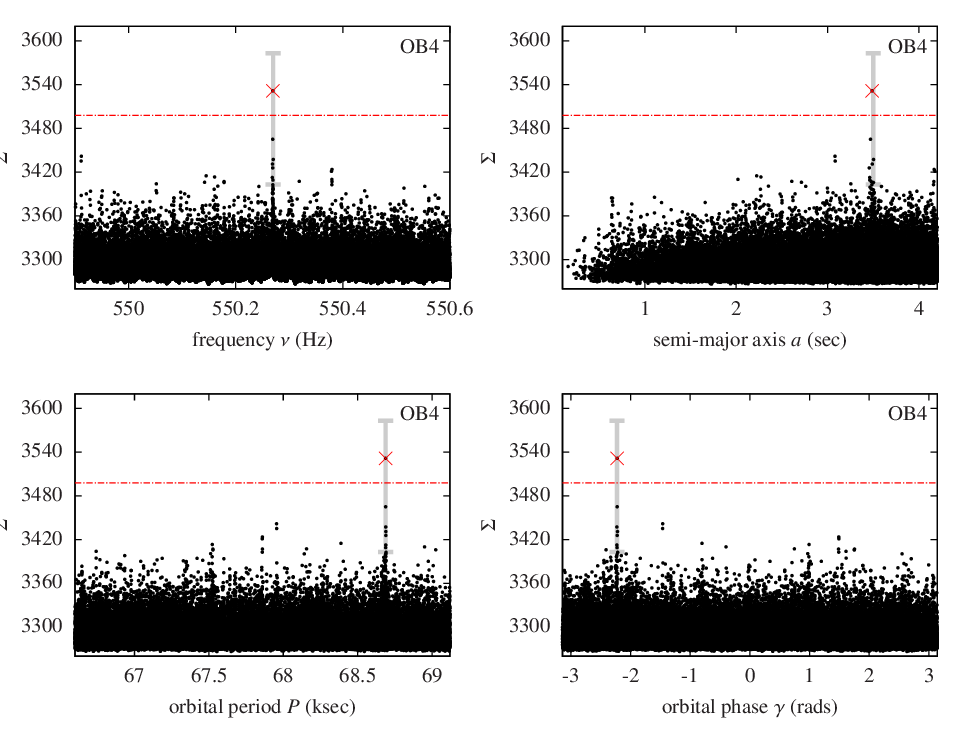}
  \end{center}                                                                  
  \caption{The semi-coherent detection statistic $\Sigma$ plotted as a
    function of the 4-dimensional physical search space for an
    artificially generated outburst based on the 4th Aql X--1 outburst
    and containing an artificial signal. The red cross indicates the
    location and value of the loudest detection statistic and the
    dashed horizontal line indicates the $1\%$ false alarm
    threshold. The y-axis span bounded by the grey lines indicate the
    1-$\sigma$ confidence region within which we expect the $\Sigma$
    value of the simulated signal's loudest template to lie.  The
    x-axis location of the grey lines indicate the parameters of the
    artificial signal.\label{fig:signal}}
\end{figure}

For the phase parameters we adopt the same approach but keep the
amplitude parameter assumed known with value $A^{*}$. The
uncertainties quoted in Table~\ref{tab:sigpars} are therefore obtained
from marginalizing the posterior distribution which in our specific
case is
\begin{equation}\label{eq:pestimation}
  p(\bm{\theta}|\Sigma^{*},\bm{\theta}^{*})\propto\chi^{2}_{2M}\left(\Sigma^{*}|,\lambda\left(A^{*},\{\mu_{\Lambda}\},\mu_{\Sigma}\left(\Delta\phi\left(\bm{\theta},\bm{\theta}^{*}\right)\right)\right)\right)
\end{equation}
where the semi-coherent mismatch $\mu_{\Sigma}$ is now expressed as a
function of the phase offsets
$\Delta\phi(\bm{\theta},\bm{\theta}^{*})$ caused by the mismatch
between the phase model of the loudest event $\bm{\theta}^{*}$ and
that of the unknown true value $\bm{\theta}$. In Fig.~\ref{fig:signal}
we show that in addition to detection, all the binary and spin
parameters were correctly recovered.
\begin{table}
  \centering
  \caption{The artificial signal parameters and their estimates.
    \label{tab:sigpars}}
  \begin{tabular}{ccccc}
    \hline \hline
    Parameter & Units & Value & Estimate & Uncertainty\\ 
    \hline
    $A$ & $\%$ & 0.40 & 0.43 & (0.39,0.46)\\
    $f$ & Hz & 550.27 & 550.26929 & 0.0013\\
    $a$ & s & 3.5 & 3.487 & 0.035\\
    $P$ & hr & 19.08 & 19.08005 & 0.00063 \\
    $\gamma$ & rads & 4.058352257914 & 4.0579 & 0.0144\\
    \hline
    \multicolumn{3}{l}{Max detection statistic $\Sigma^{*}$} & \multicolumn{2}{c}{3562.8}\\
    \multicolumn{3}{l}{Expected $68\%$ $\Sigma^{*}$ range} & \multicolumn{2}{c}{$(3403,3583)$}\\
    \multicolumn{3}{l}{Statistical significance $P_{\Sigma^{*}}$} & \multicolumn{2}{c}{$\leq 5.25\times 10^{-4}$} \\
    \hline \hline
  \end{tabular}
  \begin{tablenotes}
    \footnotesize
    \item 
 \end{tablenotes}
\end{table}
%

\section{Discussion}\label{sec:discussion}

The semi-coherent search presented in this paper represents the first
complete search of pulsations in Aql X--1, carried for all 18 outbursts
recorded with high time resolution data.  This search places strong
constrains on the presence of pulsations for 15 out of 18 outbursts,
with upper limits of ${\sim}0.3-0.9\%$. In the remaining three
outbursts the upper limits are of the order of 1--3\% (due to the
short duration of the observations).


The non-detection of pulsations in Aql X--1 adds to previous deep pulse
searches carried on 15 low mass X-ray binaries~\citep{vau94,dib05}.
\citet{vau94} analysed \textit{Ginga} data of 15 bright Z and atoll
sources with the quadratic coherence recovery technique and placed
upper limits between 0.3\% (in Sco X-1) and 8\% (4U 1608-52).
\citet{dib05} used acceleration searches on the ultra-compact source
4U 1820-30 (which was also among the 15 sources analysed by
\citealt{vau94}) and placed upper limits of the order of 0.8\% on the
pulse amplitude.

Even if many other LMXBs have never shown pulsations, the
non-detections in Aql X--1 are somehow surprising.  Indeed this source
has shown pulsations for ${\sim}150$ seconds during its 1998 outburst
(outburst 3 in table~\ref{tab:datapar}) out of a total observing time
of 1.7 Msec, so that pulsations are present in just 0.009\% of the
observed time). \citet{cas08} reported a fractional semi-amplitude of
($1.9\pm0.2)\%$ for the single pulse episode in the full \textit{RXTE}
energy band (2--60 keV). The pulse semi-amplitude in the energy band
considered here (7--25 keV) reaches a value of ${\sim}6.5\%$ whereas
our upper limits on the pulsed amplitude in the same outburst reach a
value of 0.26\%. The high pulsed amplitude of the signal makes it
difficult to believe that very weak pulses still exist and remain
undetected below our detection threshold, since this would require a
sudden jump by more than a factor of 25 in pulsed amplitude without
other pulse episodes with intermediate values being present (which we
would have detected). Our results support therefore the idea that Aql
X--1 has shown a single pulse episode.

\citet{cas08} discussed several possible scenarios to explain the
single pulse episode. In the following we review those mechanisms and
we explore new possibilities emerged in the last few years.

The presence of a dipolar magnetosphere with a field of $10^7$--$10^9$
G, comparable to that seen in radio and other AMXPs seems hard to
justify since the interaction between the field lines and the plasma
would very likely break the high degree of symmetry required to avoid
the production of pulsations. On the other hand weak pulsations are
seen in some AMXPs, most remarkably pulsations at the 0.4\% level
(0.3\% rms) where detected in the intermittent pulsar HETE
J1900.1--2455~\citep{gal08, pat12c}.  \citet{pat12c} reported that in
that particular source the pulsations are seen at the 0.4\% level only
very intermittently and suggested that this behaviour is related to
the screening of the magnetic field. In Aql X--1 such gradual screening
cannot be the explanation for the lack of pulsations because
the single 150 s is preceded and followed by the absence of
pulsations. Furthermore the magnetic field cannot re-emerge and be
screened on such short timescales, which are thought to be on the
order of the Ohmic diffusion timescale (typically 1-10
years~\citep{cum01}).

An alternative model suggests that the lack of pulsations is due to
the nearly perfect alignment of magnetic and spin axis. \citet{lam09,
  lam09b} modeled the emission of a 400 Hz AMXP, with an hot spot with
an angular size of $25^{\circ}$ and a neutron star of 1.4$\msun$ and
$10$ km in radius. According to this model, the pulse amplitude is
smaller than our most stringent upper limits only if the observer
inclination is smaller than about 10 degrees and the hot spot
misalignment angle is less than 2 degrees (see Figure 1 in both
\citealt{lam09} and~\citealt{lam09b}, with the caveat that Aql X--1 is
spinning at 550 Hz). Although initially believed to be a low
inclination binary~\citep{gar99, sha98}, Aql X--1 is now thought to
have an inclination with
$36^{\circ}<i<70^{\circ}$~\citep{wel00}.  In this case the hot spot
misalignment needs to be substantially less than 2 degrees. Since the
150 s pulse episode had a fractional amplitude of ${\sim}6\%$ (7--25
keV), which requires a misalignment of about $15^{\circ}$, it seems
difficult to conceive a mechanism to keep the hot spot almost
completely locked to the rotational axis for the greatest majority of
its lifetime and then justify a sudden large drift of $15^{\circ}$ or
more for just 150 seconds. Also, numerical MHD simulations of hot
spots on accreting neutron stars~\citep{kul13} show that such rigid
locking is nearly impossible to achieve as hot spots move and change
shape substantially during the accretion process.

Another possibility is that Aql X--1 spends most of its time accreting
via a Rayleigh-Taylor instability.  Such interchange instability has
been observed to emerge in numerical MHD simulations of
AMXPs~\citep{kul08} when the mass accretion rate overcomes a certain
threshold. Aql X--1 is the most luminous AMXP known, reaching peak
luminosities $>10^{37}\ergs$ thus indicating a high mass accretion
rate. However, it is not clear why the pulsations are not seen during
the outbursts rises, or why the single pulse episode is observed when
the luminosity has almost reached its maximum (when the mass accretion
rate is higher and thus pulsations should not be expected).

The smearing of the pulsation due to gravitational lensing is also a
possibility considered in the literature~\citep{woo88,
  oze09}. However, also in this case the presence of one single
moderately high amplitude pulsation seems to require a strong fine
tuning of the geometric configuration of the hot spot and neutron star
parameters and can almost certainly be ruled out.
Finally, smearing of pulsations via electron scattering~\citep{bra87,
  tit02} seems also difficult to justify because no spectral variation
are observed between the pulse and non-pulsating phases (see
~\citealt{cas08, alt08b} for a discussion). 

None of the mechanisms above, which do require a dipolar magnetosphere
(a multipolar magnetosphere would run into similar problems), seem to
explain the sharp contrast between the pulsating and non-pulsating
phases of Aql X--1. Although the pulse non-detections make any
scenario highly speculative, we suggest that the lack of pulsations is
related to the lack of a strong magnetosphere. We can speculate that Aql
X--1 has either no magnetosphere or a very weak one which is unable to
influence the accretion flow in any significant way. The single pulse
episode must therefore be ascribed to some other phenomenon, unrelated
to channeled accretion.

Modes of oscillations have been suggested as a possible mechanism for
the pulse episode of Aql X--1~\citep{cas08}.  An oscillation mode with
azimuthal number $m$ and frequency $\nu_{\text{mod}}$ would give an
observed frequency $\nu_{\text{obs}}$ given a spin frequency $\nu$:
\begin{equation}
\nu_{\text{obs}} = m\,\nu\,+\,\nu_{\text{mod}} 
\end{equation}
Since $\nu_{\text{obs}}=550.273$ Hz and since we know the approximate
spin frequency within ${\sim}1\,$Hz from burst
oscillations~\citep{zha98}, then an $m=1$ mode with
$\nu_{\text{mod}}{\sim}1$ Hz can explain the observations. Any shorter
mode frequency would still be a valid possibility down to a frequency
of $\nu_{\text{mod}} = 1/\Delta\,T = {\sim}7\times10^{-3}$ Hz, where
$\Delta\, T{\sim}150$ s is the duration of the single pulse episode.
However, even if the frequencies have plausible values, this mechanism
remains difficult to justify.  Indeed is not clear what
might have excited the mode since no burst or any other
relevant event has been recorded close or during the pulse episode.


\section{Conclusions}\label{sec:conclusions}

We have analyzed the entire \textit{RXTE}/PCA datasets available for
the LMXB Aql X--1 to search for pulsations with a new technique known
as semi-coherent search. We have reached an unprecedented sensitivity
that reaches a fractional amplitude of 0.3\%. We detect no
pulsations beside the already known 150-s long episode detected in
1998.  Out typical upper limits on the fractional amplitude span a
range of 0.3-0.9\% (semi-amplitude) in the 7--25 keV energy range.  By
considering all possible known pulse formation mechanisms we conclude
that Aql X--1 is unlikely to be accreting from an extended
magnetosphere and some other exotic explanation, yet to be identified,
is required to justify the observed behaviour.

\acknowledgements{A.P. acknowledges support from the Netherlands
  Organization for Scientiﬁc Research (NWO) Vidi fellowship. We would like
to thank B. Haskell and H. Pletsch for interesting discussions and suggestions.
CM would especially like to thank B. Allen and the Max-Planck-Institut f\"ur
Gravitationsphysik for their support during the initial stages of this work.}

\appendix

\section{Additional results}\label{app:results}

\begin{figure}
  \begin{center}                                                                
    \includegraphics[width=0.48\columnwidth]{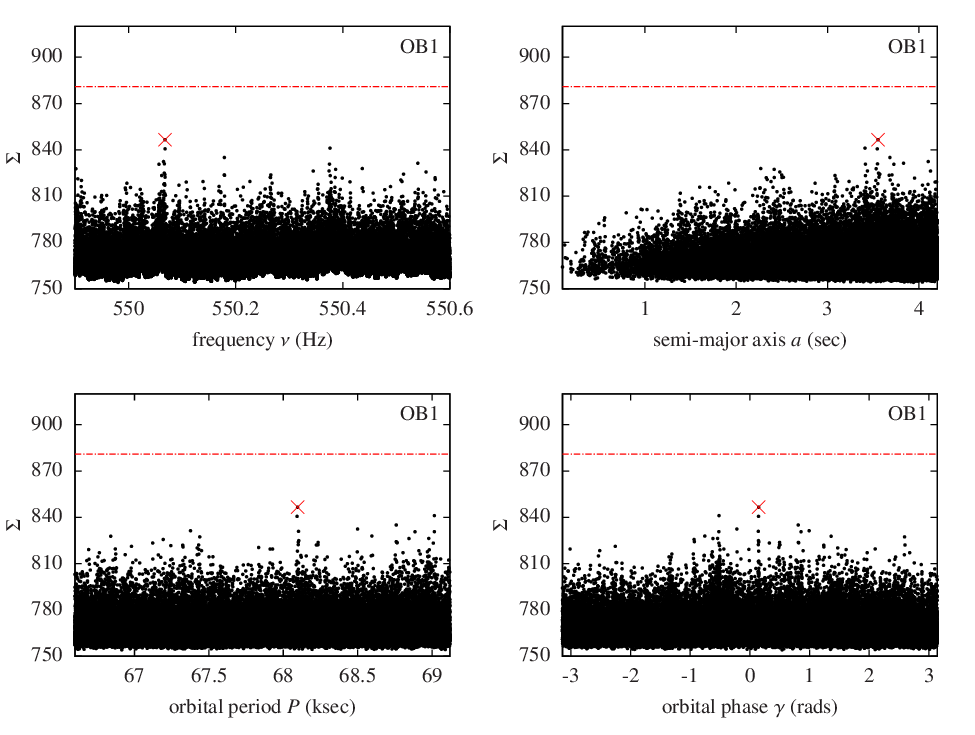}             
    \includegraphics[width=0.48\columnwidth]{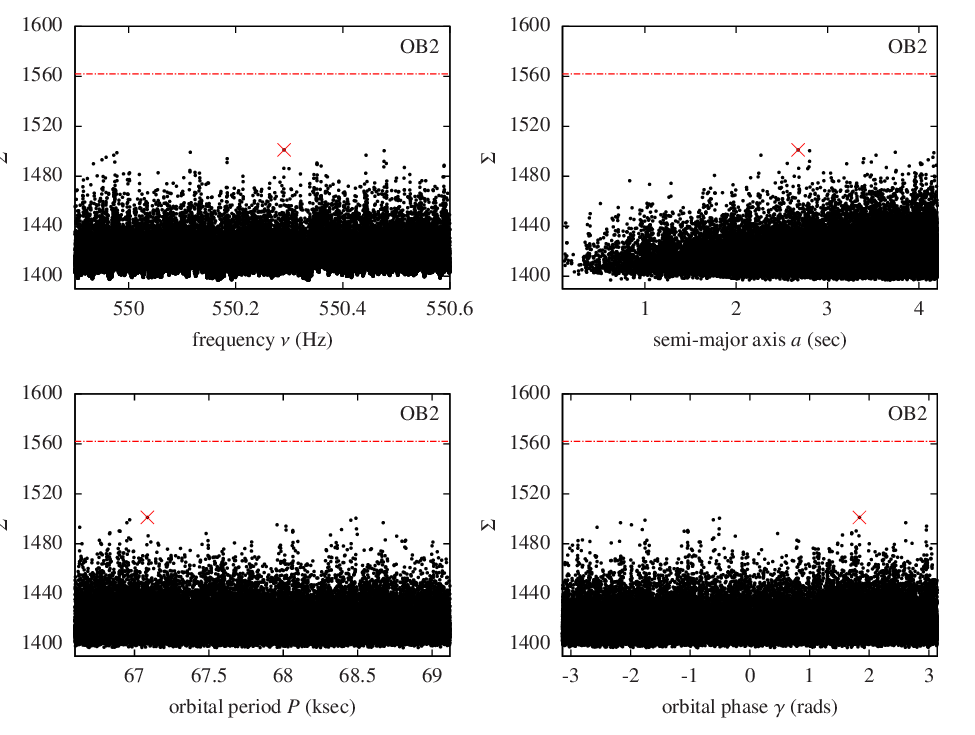}        
    \includegraphics[width=0.48\columnwidth]{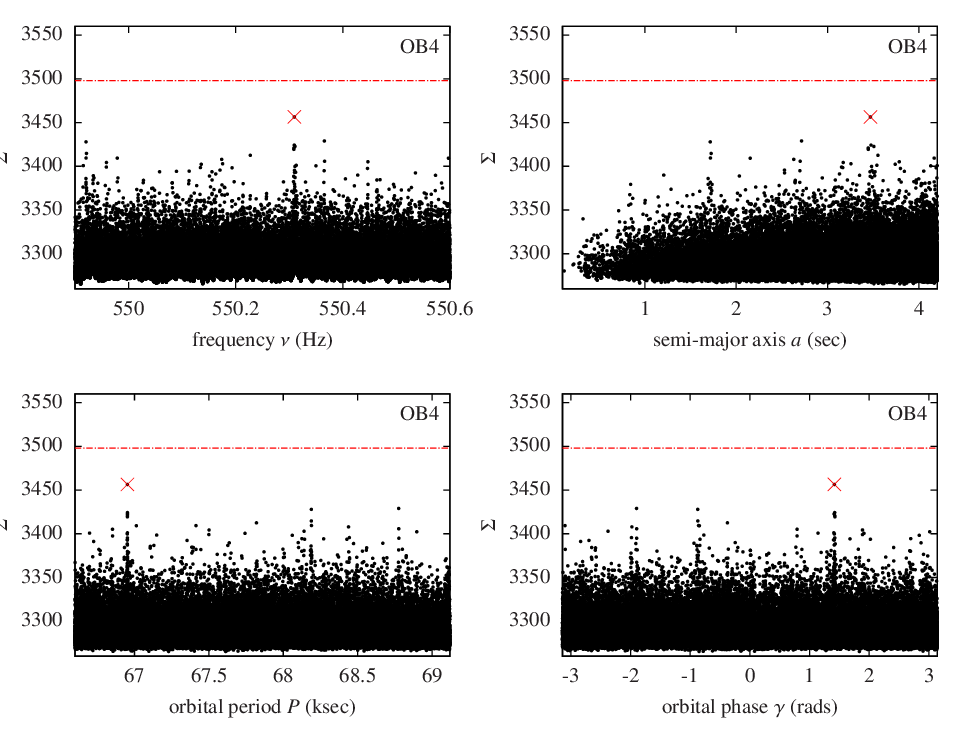}
    \includegraphics[width=0.48\columnwidth]{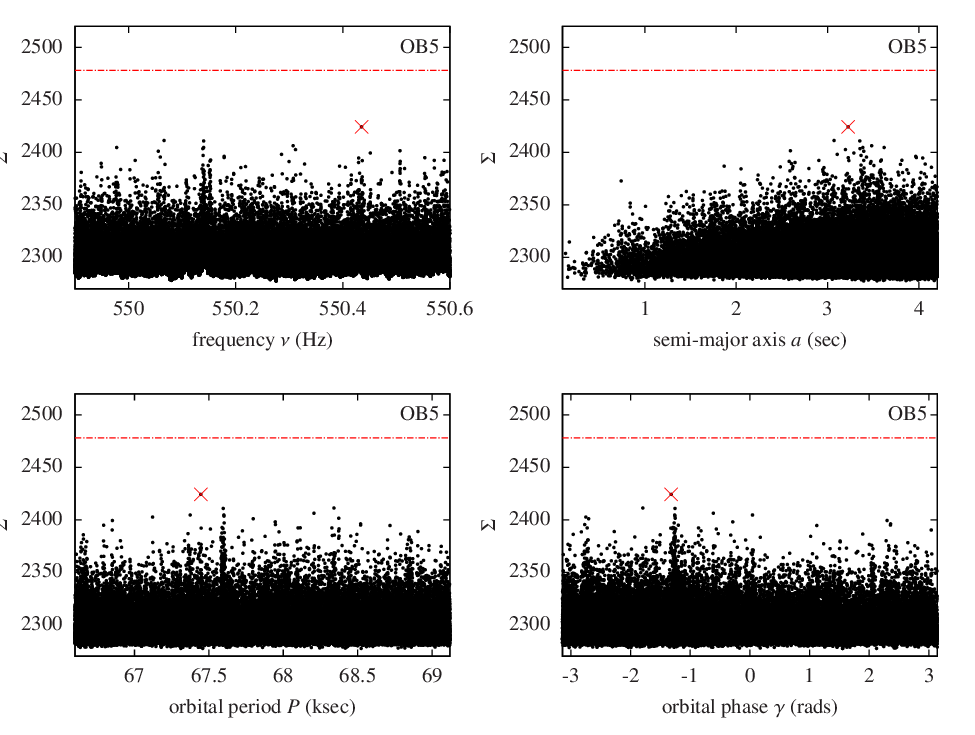}
    \includegraphics[width=0.48\columnwidth]{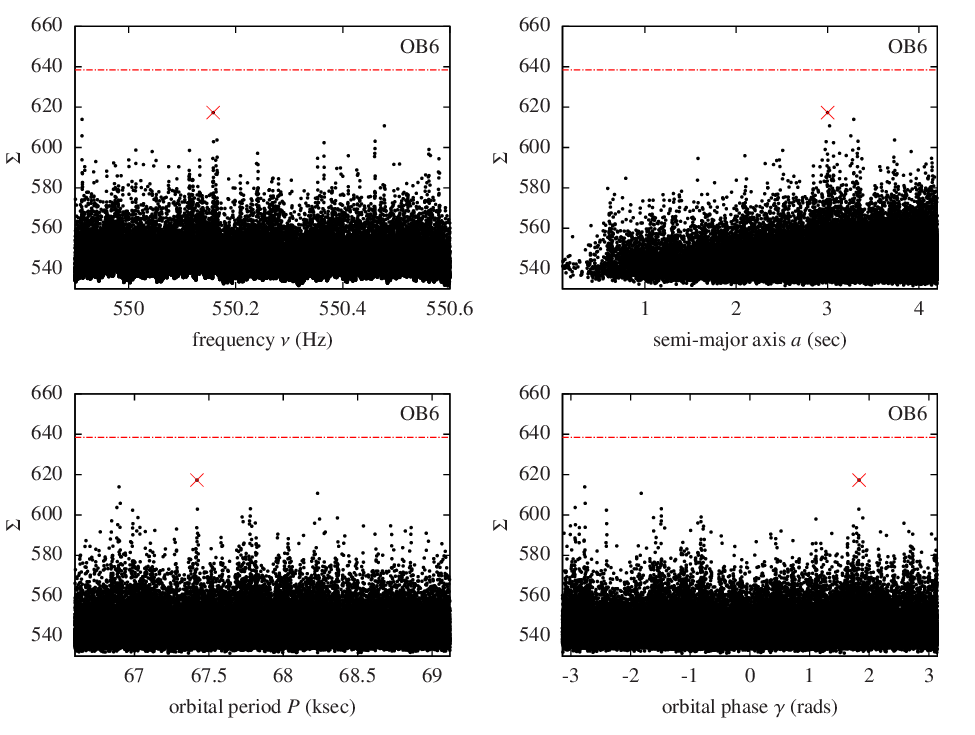}
    \includegraphics[width=0.48\columnwidth]{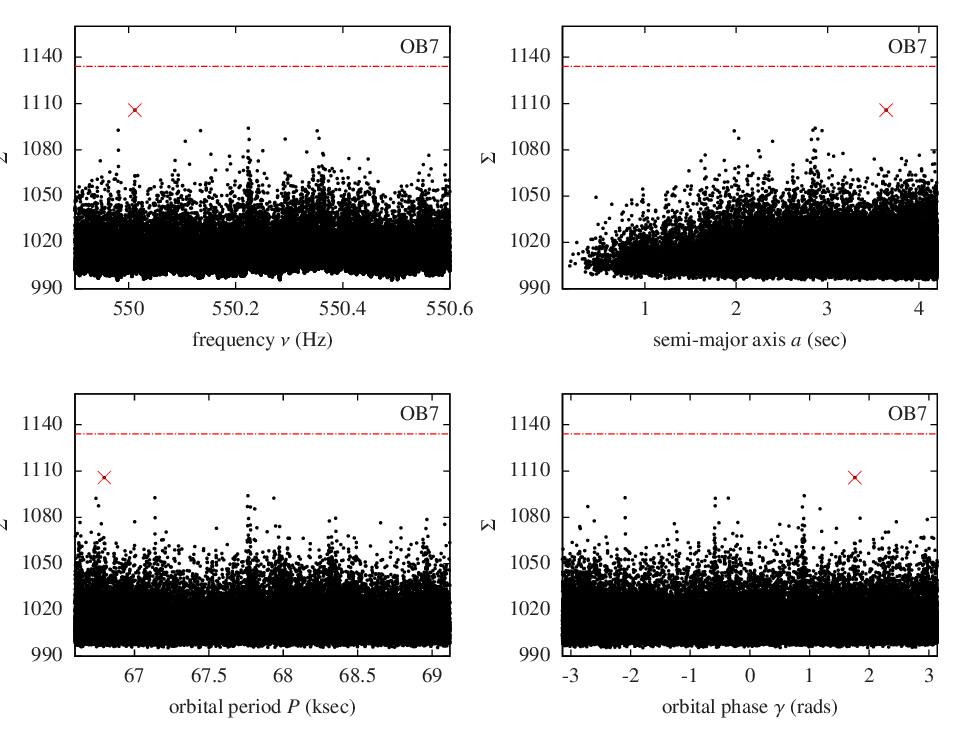}
  \end{center}                                                                  
  \caption{The semi-coherent detection statistic $\Sigma$ plotted as a
    function of the 4-dimensional physical search space for the 1st,
    2nd, 4th, 5th, 6th and 7th Aql X--1 outbursts. The red crosses
    indicates the location and value of the loudest detection
    statistic and the dashed horizontal lines indicates the $1\%$
    false alarm thresholds.}\label{fig:ob1-7}
\end{figure}
\begin{figure}
  \begin{center}                                                                
    \includegraphics[width=0.48\columnwidth]{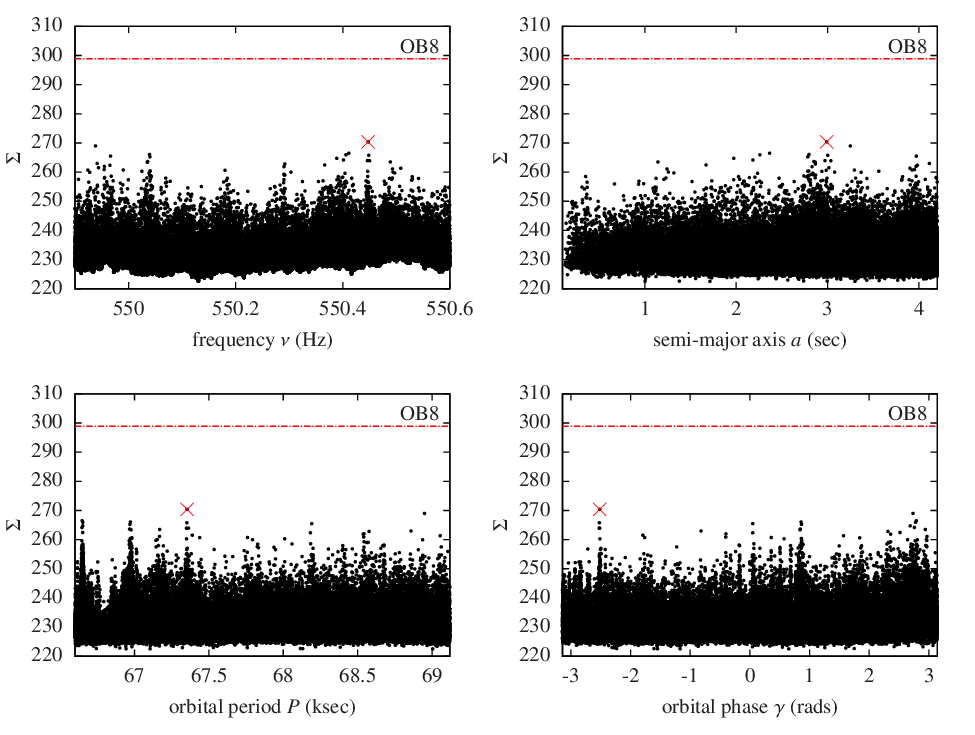}             
    \includegraphics[width=0.48\columnwidth]{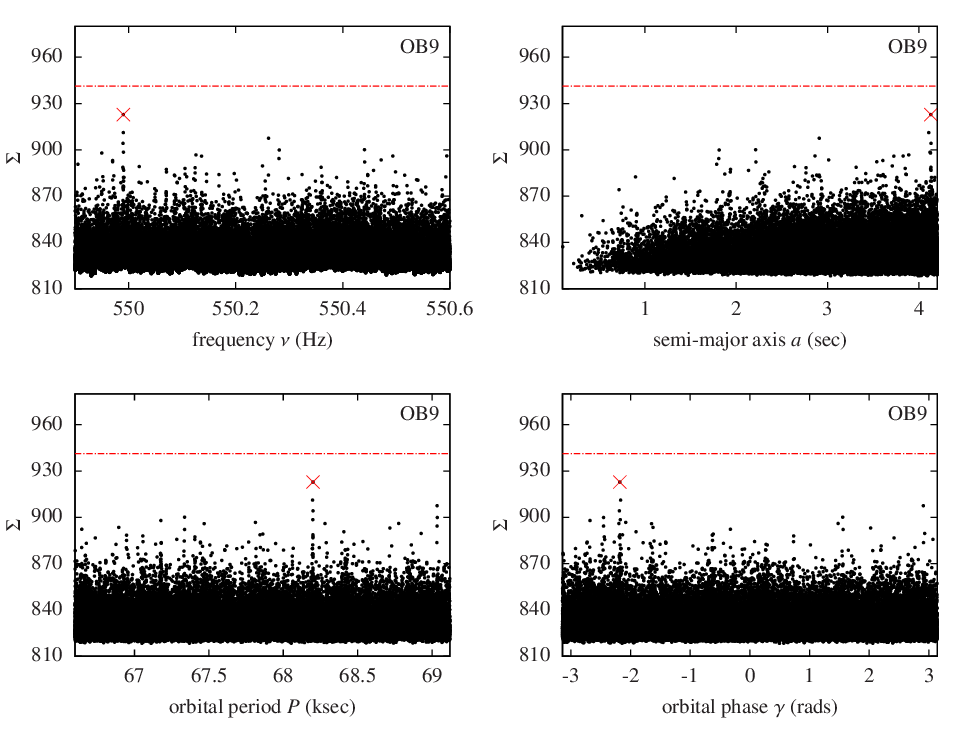}        
    \includegraphics[width=0.48\columnwidth]{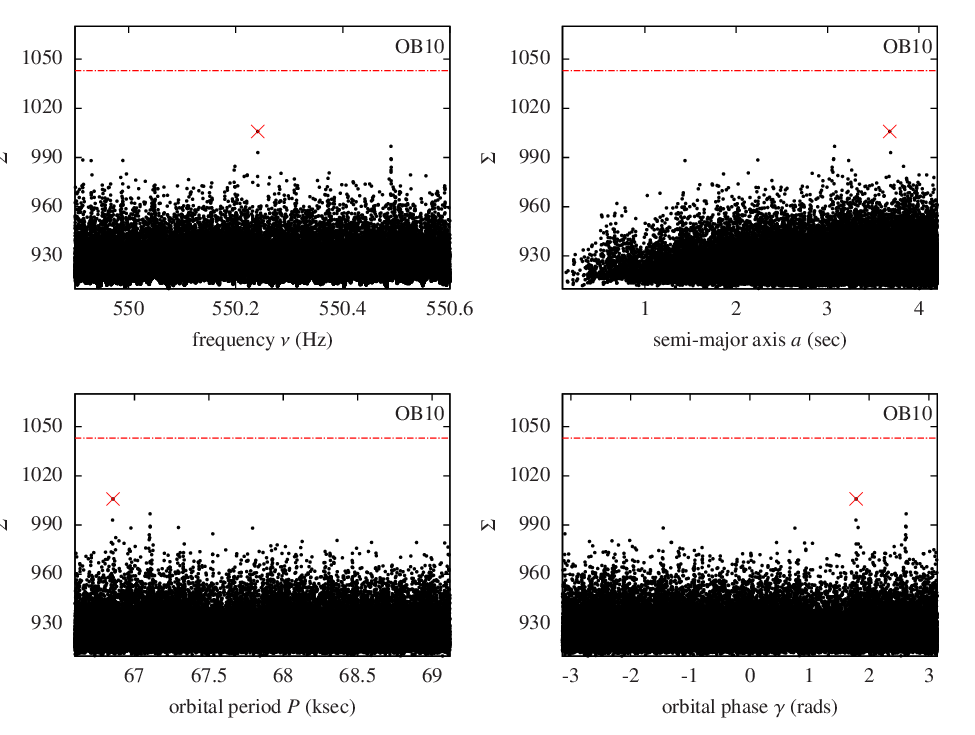}
    \includegraphics[width=0.48\columnwidth]{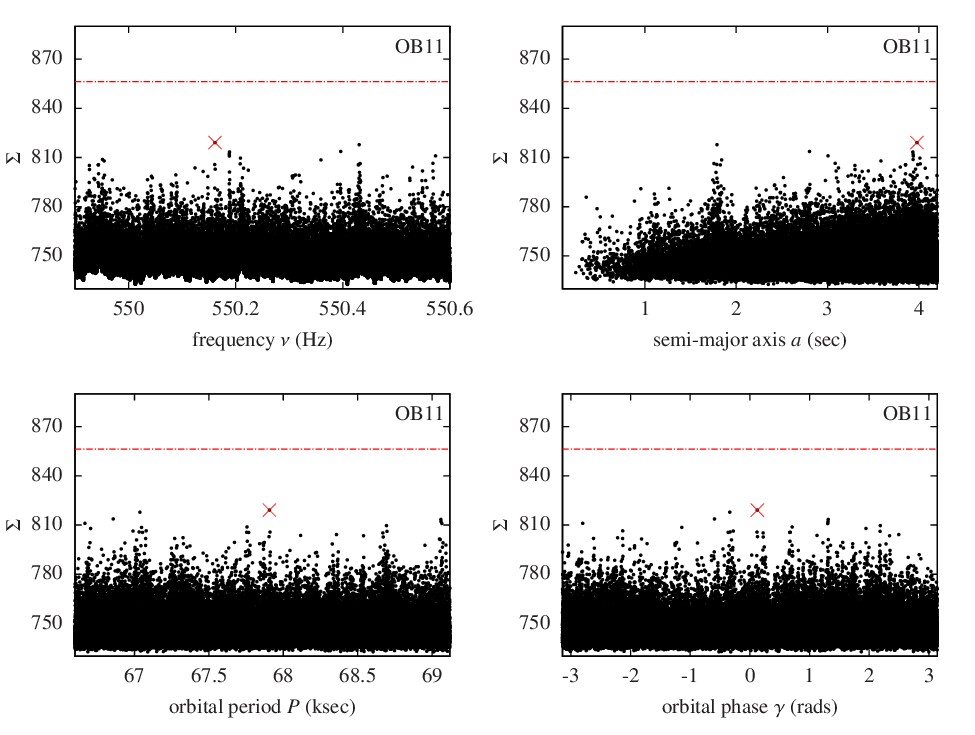}
    \includegraphics[width=0.48\columnwidth]{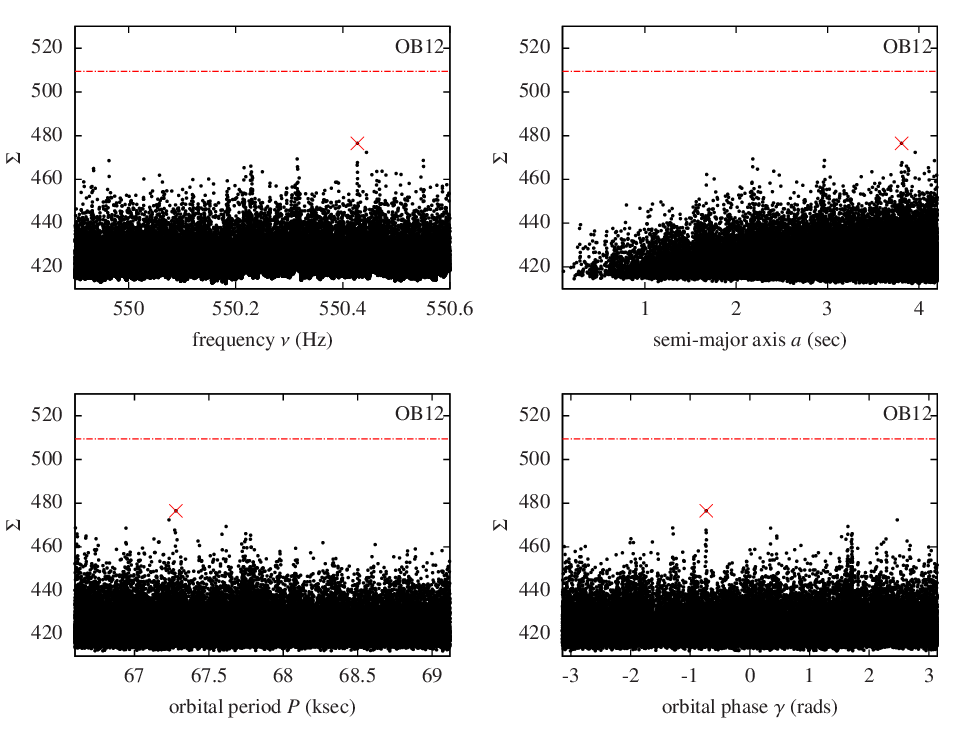}
    \includegraphics[width=0.48\columnwidth]{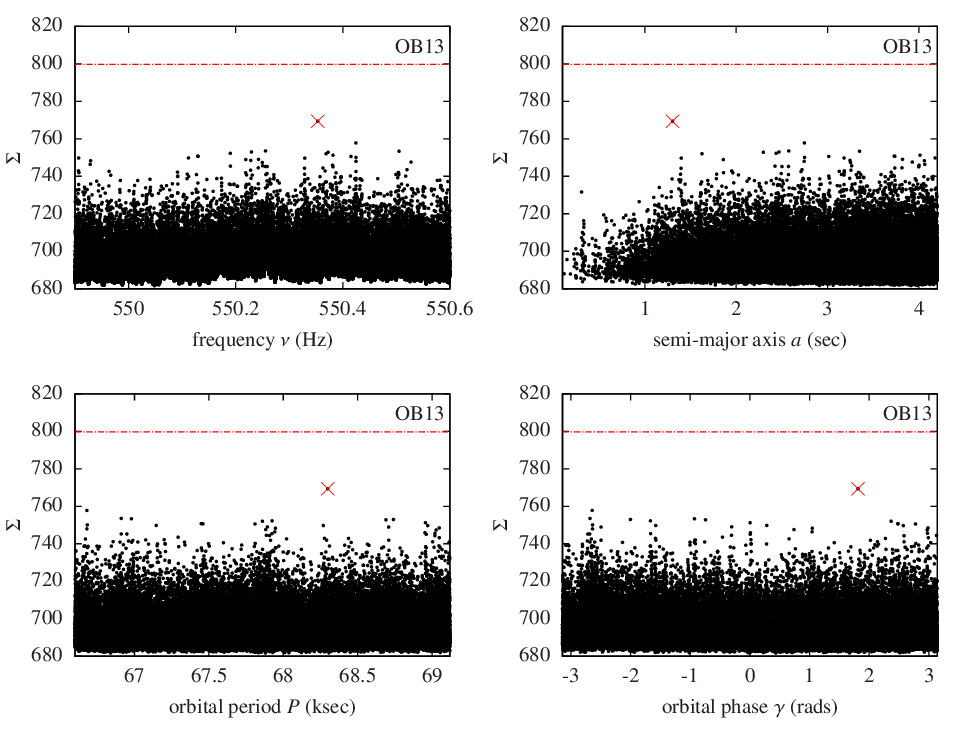}
  \end{center}                                                                  
  \caption{The semi-coherent detection statistic $\Sigma$ plotted as a
    function of the 4-dimensional physical search space for the
    8th--13th Aql X--1 outbursts. The red crosses indicates the
    location and value of the loudest detection statistic and the
    dashed horizontal lines indicates the $1\%$ false alarm
    thresholds.}\label{fig:ob8-13}
\end{figure}
\begin{figure}
  \begin{center}                                                                
    \includegraphics[width=0.48\columnwidth]{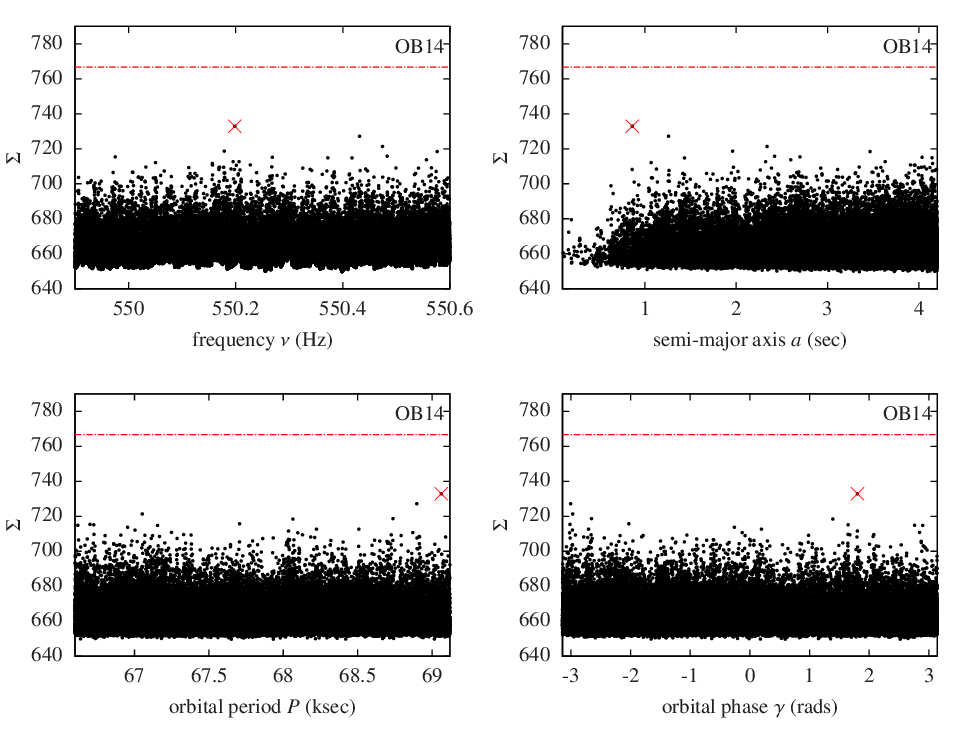}             
    \includegraphics[width=0.48\columnwidth]{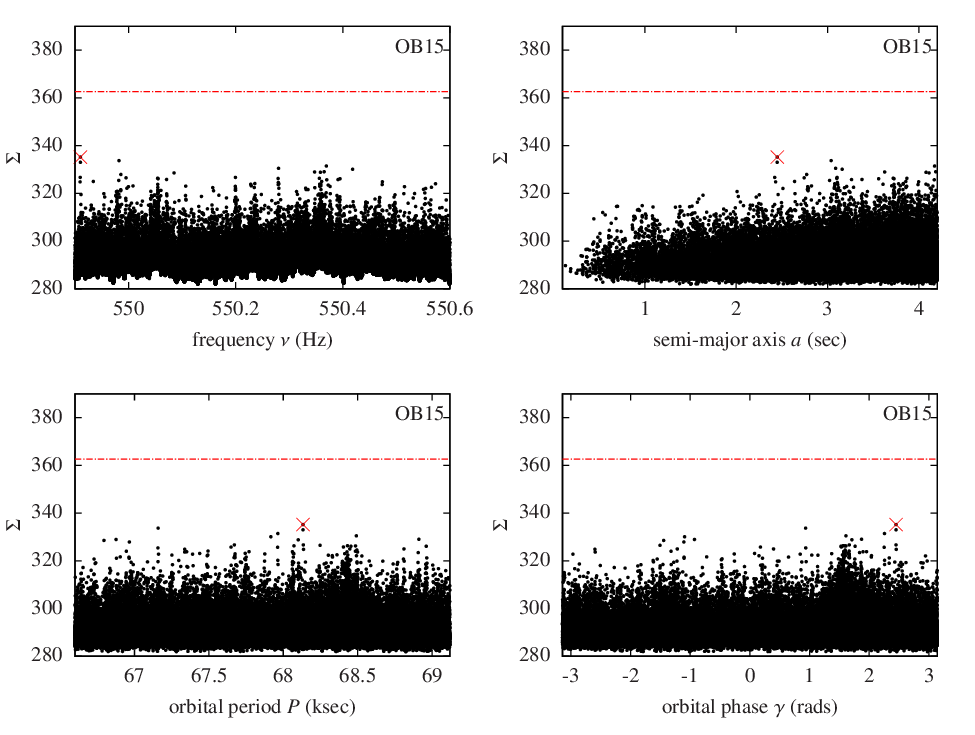}        
    \includegraphics[width=0.48\columnwidth]{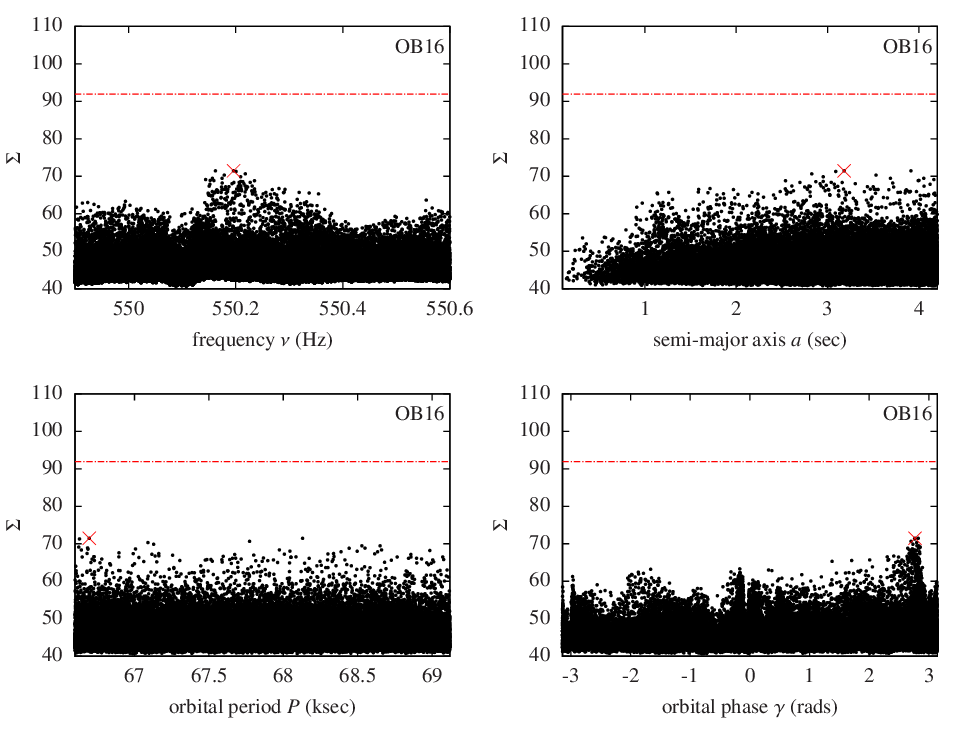}
    \includegraphics[width=0.48\columnwidth]{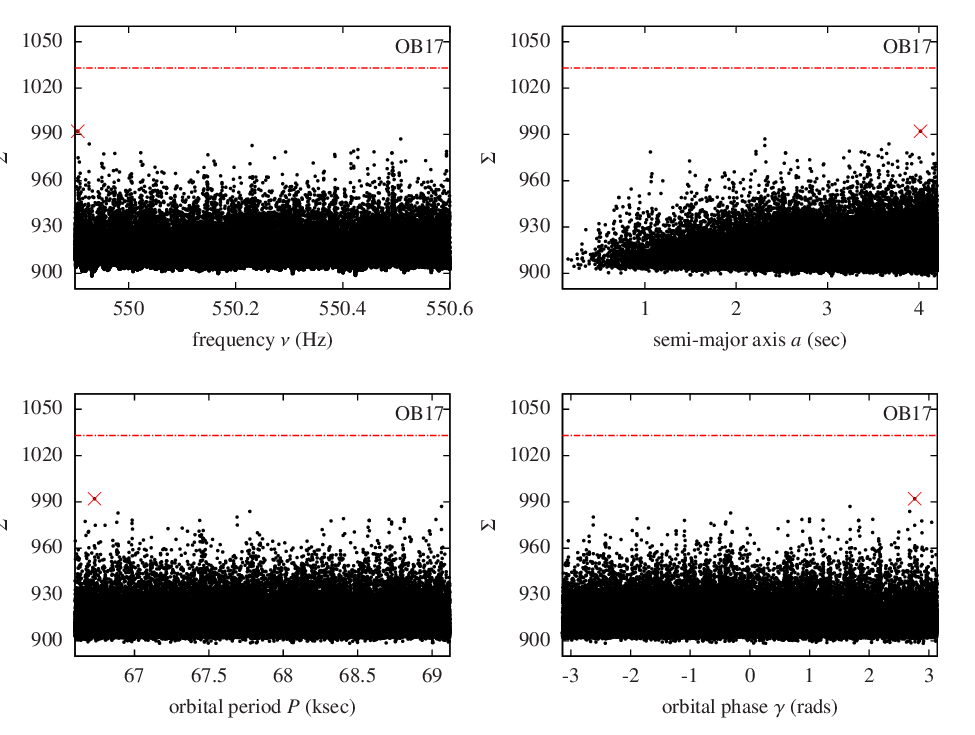}
    \includegraphics[width=0.48\columnwidth]{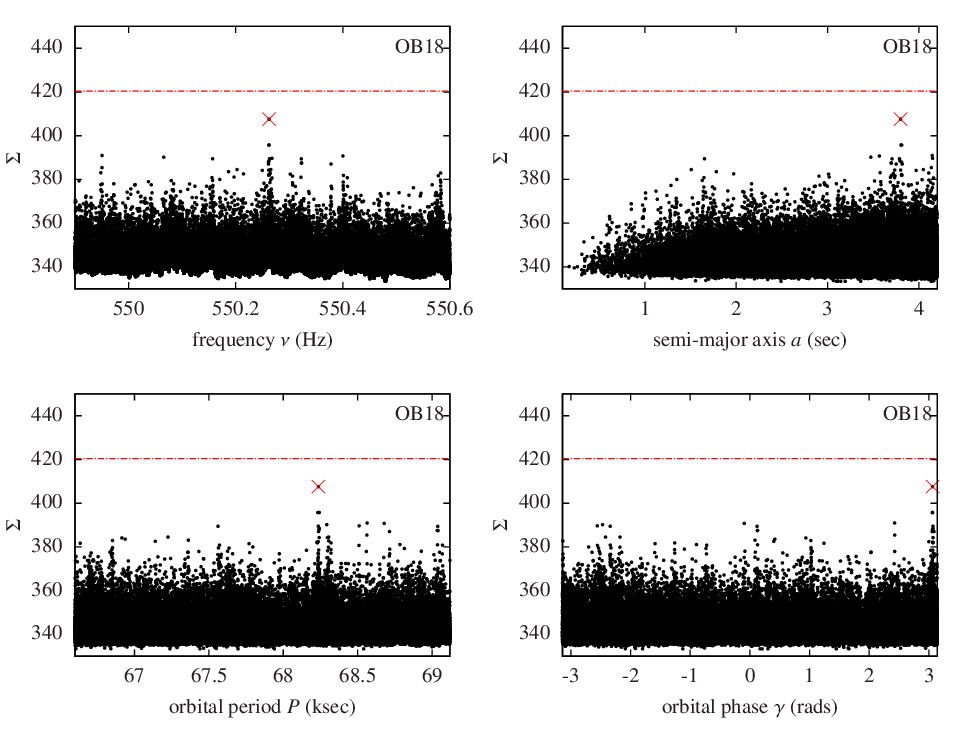}
  \end{center}                                                                  
  \caption{The semi-coherent detection statistic $\Sigma$ plotted as a
    function of the 4-dimensional physical search space for the
    14th--18th Aql X--1 outbursts. The red crosses indicates the
    location and value of the loudest detection statistic and the
    dashed horizontal lines indicates the $1\%$ false alarm
    thresholds.}\label{fig:ob14-18}
\end{figure}

\bibliographystyle{apj}
\bibliography{biblio}
\end{document}